\documentclass[mathpazo]{cicp}

\usepackage{geometry}

\usepackage{amsmath, amssymb, amsfonts,accents}

\usepackage{color}
\usepackage{multirow}
\usepackage{subfigure}

\usepackage{threeparttable}

\usepackage{bm}
\usepackage[normalem]{ulem}
\usepackage{todonotes}

\usepackage{algorithm}
\usepackage{algorithmic}
\usepackage{cancel}
\usepackage{epstopdf}

\usepackage{hyperref}

\usepackage{lineno,empheq}
\usepackage{eqnarray}

\newcommand{\abs}[1]{\left\vert#1\right\vert}
\newcommand{\set}[1]{\left\{#1\right\}}
\newcommand{\Real}{\mathbb{R}}

\begin{document}

\title{An Improved Adaptive Minimum Action Method for the Calculation of Transition Path in Non-gradient Systems}

\author[Sun Y Q et.~ al]{Yiqun Sun\corrauth, Xiang Zhou}
\address{Department of Mathematics,\\ City University of Hong Kong\\
Tat Chee Ave, Hong Kong SAR}
\emails{{\tt yqsun5-c@my.cityu.edu.hk} (Y.~Sun), {\tt xiang.zhou@cityu.edu.hk} (X.~Zhou)}

\begin{abstract}
The minimum action method (MAM) is to calculate the most probable transition path in
randomly perturbed stochastic dynamics, based on the idea of action minimization in the path space. The accuracy of the numerical path between different metastable states usually suffers from the ``clustering problem'' near fixed points. The adaptive minimum action method (aMAM) solves this problem by relocating image points equally along arc-length with the help of moving mesh strategy. However, when the time interval is large, the images on the path may still be locally trapped around the transition state in a tangle, due to the singularity of the relationship between arc-length and time at the transition state. Additionally, in most non-gradient dynamics, the tangent direction of the path is not continuous at the transition state so that a geometric corner forms, which brings extra challenges  for the aMAM. In this note, we improve the aMAM by proposing a better monitor function that does not contain the numerical approximation of derivatives, and taking use of a generalized scheme of the Euler-Lagrange equation to solve the minimization problem, so that both the path-tangling problem and the non-smoothness in parametrizing the curve do not exist. To further improve the
accuracy, we  apply the Weighted Essentially non-oscillatory (WENO) method for the interpolation to achieve better performance. Numerical examples are presented to demonstrate the advantages of our new method.

\end{abstract}

\ams{82C26, 60H30, 34F05}
\keywords{Rare Event, Transition Path,  Minimum Action Method, Moving Mesh, WENO}

 \maketitle

\section{Introduction}
\label{sec1}
The calculation of quasi-potential and the most probably transition path between stable equilibria in metastable systems is of interest to researchers in the study of dynamics of complex and stochastic systems in long time scales \cite{FW2012}. The minimum action method (MAM) \cite{weinan-MAM2004} was   introduced to find such optimal paths by directly minimizing the action functional, which is the rate function in the large deviation theory \cite{FW2012}. The action minimizer within a certain rare-event subset in the path space, i.e., the minimum action path, carries the dominant contribution  to the probability for the corresponding rare events. So, conditioned on the occurrence of rare events, the minimum action path is the most probable path under the influence of small noise in the long run. In many applications, the path represents  the progressive physical process of important rare events such as phase transformation, chemical reaction, etc. Therefore, the numerical study of how to calculate the path efficiently is of great importance.

There have been quite many developments of numerical methods for the minimum action path. Firstly, when the system is of gradient type, {\it i.e.}, the dynamics is gradient flow driven by a potential energy, the variational problem for the minimum action path gets simplified and it turns out that essentially the  optimal transition path is simply the time-reversed trajectory of the gradient flow. The joint location of  the ``uphill'' path from one well and the ``downhill'' path toward the other well in the phase space is an index-1 saddle point which serves as transition state. The min-mode eigen-direction of the saddle point collapses with the tangential direction of the path from the {\it both} sides. This result reveals a second important feature for the gradient system: the tangential direction of the path is always continuous (belonging to $C^1$ curve), even when it crosses the separatrix via the saddle point. In practice,  the path-finding algorithms, such as the string method \cite{String2002} never searches the ``uphill'' and ``downhill''  paths separately since the saddle point is unknown {\it a priori}, but these methods search the whole path once for all between local minima and then locate the saddle point from the numerical path and split the path into ``uphill'' and ``downhill''  segments for interpretations.

But in the non-gradient systems,  the above features of the path no longer holds, due to the lack of detailed balance for the stationary probability distribution. The ``uphill'' path  and the ``downhill'' path are distinctively different in nature.  More importantly, they meet at the saddle point from different directions in two sides of the separatrix:  a sharp corner is usually  formed where the path crosses the separatrix.  The ``uphill'' path escapes the characteristic  boundary by choosing a direction  different from the  eigen-direction of the saddle point. The path then exhibits non-smoothness,  meaning that  if the path $\varphi$ is written in terms of arc length parameter $s\in[0,1]$,  then the tangent vector  $\varphi'(s)$ is not continuous at  $s_*$, even though $\abs{\varphi'(s)}\equiv const$, where $\varphi(s_*)$ is the location of  the saddle point.   This is a generic phenomenon for transition path in non-gradient system and it is the origin of the non-Gaussian skewed distribution of the exist point on boundary. It also affects  the prefactor estimation of mean exist time  for non-gradient systems \cite{MatSchuss1983,Maier1992PRL}.  Refer to the work of \cite{Maier1993PRE,Maier1993PRL,Stein2016PRE}  for theoretical analysis of the connection to effect of focusing and caustics. To visualize this non-smooth feature of the paths, we shall present two examples including the  Maier-Stein model  in \cite{Maier1996JSP}  in the figures shown later.

To locate the saddle points,  on the other hand, the direct search of saddle point, such as the dimer method \cite{Dimer1999} or the gentlest ascent dynamics\cite{GAD2011,SamantaGAD}, sounds an alternative strategy. Actually,  in many existing  practical applications \cite{Lorenz2008, {KS-WZE2009},XWan2013JCP}, after applying the standard MAM, one  can  usually obtain a reasonable result of the path and then by checking the force along the path, one can further  guess the locations of critical points along the curve to serve as initial guess  for a further independent run of  saddle-point search algorithms  (from the most straightforward  Newton method to  the new method of GAD \cite{GAD2011}).  After successfully splitting the path into ``uphill" and ``downhill" pieces ,  one can then run the MAM again separately  for each   piece to  further refine the path and the quasi-potential.   This approach only works when the  numerical path in the first round of calculation is not too far away from the true solution  and also critically relies on the success of finding saddle points. So it should be regarded as a post-process step rather than a true path-finding approach.

It seems there has  no much work of a  natural and  accurate numerical method  to directly address the issue of non-smoothness of the path, without extra  interference or   post-processes mentioned above. The discontinuous tangential direction across the boundary arising in non-gradient systems imposes very subtle challenges for the numerical calculations of  the transition path. In the    Freidlin-Wentzel theory, the action functional is  defined  in the absolute continuous function space  $AC[0,T]$ for a fixed $T>0$, while the optimal $T$ that further minimizes the action is infinitely. Actually, at the saddle point, say $x_*$, the uphill path, denoted by $\varphi_+(t)$, goes to $x_*$ as $t\to +\infty$ while the downhill path, denoted by $\varphi_-(t)$, emits from $x_*$ as $t\to-\infty$. The time derivatives $\dot{\varphi}_+$ and $\dot{\varphi}_-$ both vanish to zero as $t$ tends to $+\infty$ and $-\infty$ respectively.  In practice, the two jointly infinite-time intervals   are approximated by a single finite time interval $[0,T]$ for a very large $T$. When $T$ is large, the problem actually becomes ill-posed:  the grid points are extremely dense in geometry near fixed points. Numerous computational results of this clustering effect show the critical importance of calculating the path by adopting the arc-length parametrization. To capture the geometric   shape of the path efficiently, it is very important to represent the path by the grid points with equal geometric distance along the curve. The adaptive MAM, ``aMAM", \cite{aMAM2008} and the geometric MAM, ``gMAM", \cite{Heymann2006} were developed for this purpose by   resolving the arc-length parametrization  either  numerically or analytically. The numerical approach in aMAM  is based on the moving mesh strategy by choosing a monitor function equivalent to the derivative in time variable. The analytical approach in gMAM is based on the reformulation of the Freidlin-Wentzell action functional by using the Maupertuis' principle. The main difference between these two methods is that  aMAM works for arbitrary  but finite $T$, but gMAM   works for the optimal case $T=\infty$. The subsequent research work to improve the numerical efficiency of the MAM includes  \cite{MAM2011JCP, XWan2013JCP,XWan2016}, which covers the topics of high order time discretization and spatial discretization for applications to Navier-Stokes equation.

 As we said,  the dynamics is simply zero at the saddle points, where the non-smooth corners of the path lie.  So, one can find that the  coefficient for the second spatial derivative  in the associated Euler-Lagrange equation written in form of arc-length parametrization has a degeneracy at these zero-dynamics locations. This is the underlying reason why the a second-order equation will give arise to the non-smooth solution.  But numerically, this  actually might  not be a severe challenge because we empirically find that the simple finite-difference scheme  in the aMAM  \cite{aMAM2008}  can also guarantee the second order convergence rate, in terms of the grid size, for the $L_2$ error of the path. The real challenges in practice for the version of the aMAM proposed in \cite{aMAM2008} are:   (1) for a large time interval, {\it i.e.}, a large  domain  of $t$, the tangling phenomena usually come up and the effect for the  moving mesh breaks down. (2) the cubic spline interpolation used in the reparametrization step lowers its accuracy around the corner. In short, the scheme used in \cite{aMAM2008} is not efficient  to address both issues when a very large domain $[0,T]$ is used.

The tangling phenomena, illustrated in Figure \ref{f4} and \ref{f6} below, mean a lot of points (images)   on the path are in a tangle near the fixed point, which usually arise for a large $T$ and does not appear for the small $T$ or a small number of discrete points representing the whole path. This  implies that the moving mesh strategy failed to take effect near the saddle points and the reason is related to the monitor function used in  \cite{aMAM2008}: $\omega(t) = \abs{\varphi_t}$,  for which  the derivative  ${\varphi_t} \approx  (\varphi_{n+1}-\varphi_{n-1})/2\Delta t$ has to be calculated numerically. Near the saddle point,  the condition number in the optimization problem is very large, the images $\{\varphi_n\}$ there   behave  nearly randomly so that   any numerical  derivatives based on these ``tangled" images generates a huge error in approximating the true value, failing to correct the  random-like distribution of the images back into order. The larger domain for $t$, the higher possibility for this ``chaotic'' tangling to occur.

Here  we want to propose a new  form of the monitor function $\omega$  for the aMAM to use, so that the $L_\infty$ norm of the error in numerical path can be controlled with good accuracy. In principle, one could  introduce the idea like ENO or WENO (Weighted Essentially non-oscillatory) to approximate the derivative used by   $\omega(t)$ in a better way. However, our idea is even simpler and motivated by the heuristic and the deep connection to the underlying Hamiltonian dynamics governing the path (which is a key ingredient in gMAM). Our new monitor function only depends on the dynamics at each images --- no any derivative is involved, and has a scaling parameter $0<r\leq 1$ to further control the parametrizaton. It can also be shown that at the limit $T\to \infty$, the parametrization  corresponding to the new monitor function with  $r=1$  is still in arc-length parameter.  In this way, the tangling problem occurred in numerical solution of aMAM is avoided and the moving mesh strategy becomes more robust. The second important consequence of our new monitor function is that with the proper choice of $r<1$ in our method, the path associated with the corresponding parametrization,  $\varphi(\alpha)$ for  $\alpha\in [0,1]$ , actually does lie in $C^1([0,1])$: the singularity is removed into the relation between the arc-length and our new parametrization, not in the path-finding algorithm (see Figure \ref{fig:rsmooth} in Section \ref{sec3} later for this point). To further improve the interpolation accuracy, we also make use of the well-known  WENO method to do reparameterization, which has been widely applied in capturing the discontinuous solutions such as shock waves and imaging sciences  \cite{WENO1994JCP, {ChiwangShu1995}, {WENO1998Springer},  {WENO1999ESAIM}, {WENO1999JCP}, {ChiwangShu2002JCP},WENO2004JCP}. Interestingly, this idea of using the  WENO scheme for path calculation has already been   applied in  the string method for gradient systems  \cite{string2003CMS} to construct the higher order  scheme to improve the discretizing accuracy in handling the Euler-Lagrange equation.  However, as we already emphasized, only the path in non-gradient systems has the non-smooth structure and calls for the application of  the WENO  purely for the purpose of reconstruction of more accurately reparametrized  paths.  The traditional cubic spline interpolation for the gradient system used in \cite{string2003CMS}  is not desired for  our purpose here. We prefer  the use of the high order WENO interpolation since it works very well  regardless of the regularity of the path (as a function of $\alpha$).

The rest of the paper is organized as follows. In  Section 2, we review the minimum action method and adaptive minimum action method.  And in Section 3, we give a derivation of our improved scheme of adaptive minimum action method, and we review the WENO scheme for estimating function values in reparameterization step, then we present our  improved numerical scheme. In  Section 4,  we show numerical results of the tunnel-diode model and  the Maier-Stein model. The last is the conclusion session.

\section{Theoretical background and the Adaptive Minimum Action Method}
\label{sec2}
In this section we describe the minimum action method \cite{weinan-MAM2004} and adaptive minimum action method \cite{aMAM2008} for the calculation of minimum action path.

\subsection{Theoretical background}
Consider the system modeled by
the following It$\hat{o}$ stochastic differential equation,
\begin{equation}\label{q1}
\dot{\varphi}(t)=b(\varphi(t))+\sqrt{\epsilon}\sigma(\varphi(t))\xi(t),
\end{equation}
where $b$ is the deterministic drift field, $\sigma$ is the noise amplitude tensor, and $\xi$ is a Gaussian white noise with zero mean and covariance $\langle \xi_i(t)\xi_j(t')\rangle=\delta_{ij}\delta(t-t')$. In the framework of the large deviation (cf. \cite{FW2012}), for an SDE like (\ref{q1}) in the vanishing noise limit $\epsilon \to 0$, the probability of a path starting from $\phi_0$ at time $0$ and end in $A$ at time $T$ is given by
\begin{equation}\label{q8}
\lim_{\epsilon\downarrow0}\epsilon\log P(X_T\in A)=-\inf_{\mbox{
\tiny
$\begin{array}{c}
\varphi(0)=\phi_0,\\
\varphi(T)\in A\end{array}$
}
} S_T(\varphi),
\end{equation}
where the action functional
\begin{equation} \label{S}
 S_T(\varphi) = \frac{1}{2}\int_0^T|\dot{\varphi}-b(\varphi(t))|^2_a dt,
\end{equation}
and $a(x):=\sigma(x)\sigma(x)^T$ is the diffusion tensor, $\langle u,v\rangle_a=\langle u,a^{-1}v\rangle$ and $|u|_a=\sqrt{\langle u,u\rangle_a}$. Also, for generalization of time $T$ goes to infinity, the quasi-potential from point $\phi_0$ to point $\phi_1$ is defined by
\begin{eqnarray}\label{q4}
V(\phi_0,\ \phi_1)&=&\inf_{\tiny{T>0}}\inf_{\tiny{\varphi \in {AC}_{\phi_0}^{\phi_1}(0,T)} }S_T(\varphi),
\end{eqnarray}
where ${AC}_{x}^{y}(0,T)$ is the space of absolutely continuous functions $f: [0,T]\rightarrow R^n$ with $f(0)=x$ and $f(T)=y$. All the starting points $\phi_0$ and the ending points $\phi_1$ in concern throughout this article are two stable fixed points of the field $b(\varphi)$. And the transition rates between two states $\phi_0$, $\phi_1$ are described by the following equations
\begin{equation}\label{q11}
k_{0\rightarrow 1}\asymp \exp (-\epsilon^{-1}V(\phi_0,\ \phi_1)), \  \ k_{1\rightarrow 0}\asymp \exp (-\epsilon^{-1}V(\phi_1,\ \phi_0)).
\end{equation}
To find the minimum of action $S_T(\varphi)$ is essential in calculating probability of the event $(X_T\in A)$ in equation (\ref{q8}) and transition rate in (\ref{q11}). The Minimum Action Method calculates the following minimization problem:
\begin{equation}\label{q2}
\inf_{\tiny{\varphi \in {AC}_{\phi_0}^{\phi_1}(0,T)}}S_T(\varphi).
\end{equation}
And the minimizer $\varphi^*$ is the minimum action path (MAP), which characterize the most probable path from $\phi_0$ to $\phi_1$ under the influence of small noise. We can obtain the minimum action path by optimization algorithms to solve the problem (\ref{q2}).  To solve the minimization problem (\ref{q2}) in practice, one usually uses the standard optimization method such as  evolving the gradient flow $-\delta S/\delta \varphi$, whose steady state equation is  the Euler-Lagrange equation of the following boundary value problem
\begin{equation}
\begin{cases}\label{q23}
\varphi_{tt}- (\nabla b(\varphi)-(\nabla b(\varphi))^T)\varphi_{t}-(\nabla b(\varphi))^Tb(\varphi)=0,~~~t\in[0,T] \\
\varphi(0)  =  \phi_0,\ \ \ \varphi(T)=\phi_{1}.
\end{cases}
\end{equation}

\begin{figure}[!htbp]
\centering
\subfigure[$T=30$]{
\begin{minipage}[b]{0.45\textwidth}
\includegraphics[width=1.0\textwidth]{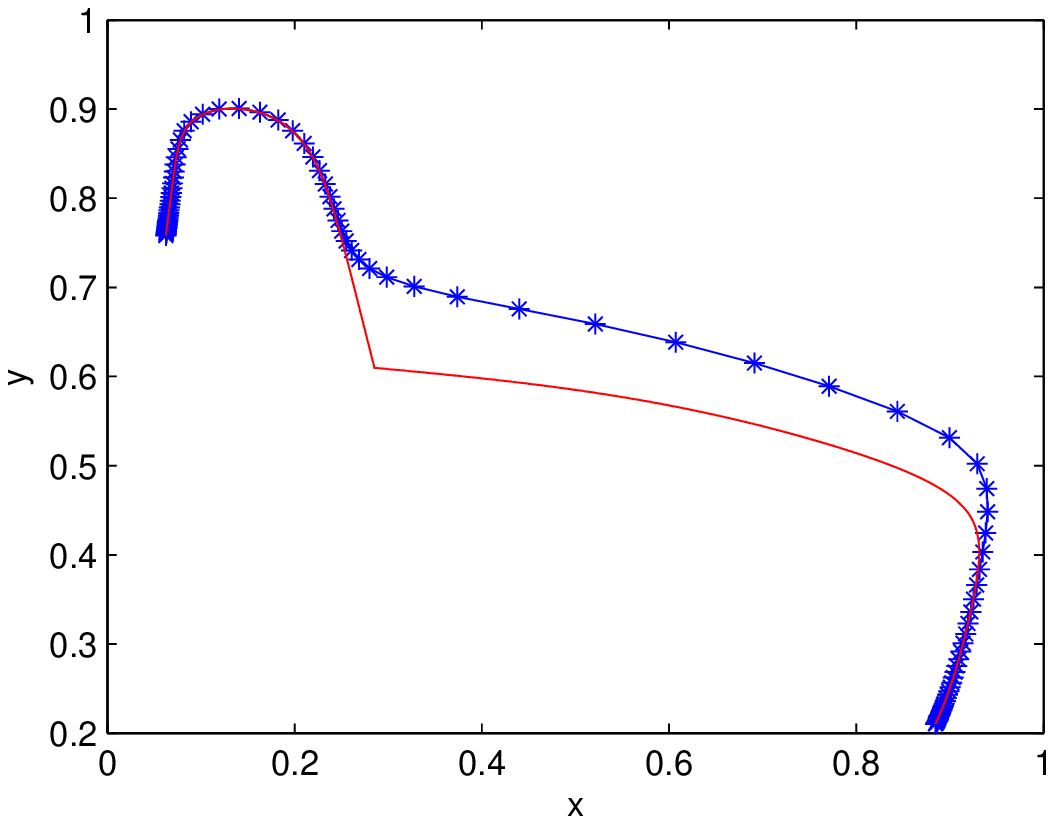}
\end{minipage}
\label{f4a}
}
\subfigure[$T=200$]{
\begin{minipage}[b]{0.45\textwidth}
\includegraphics[width=1.0\textwidth]{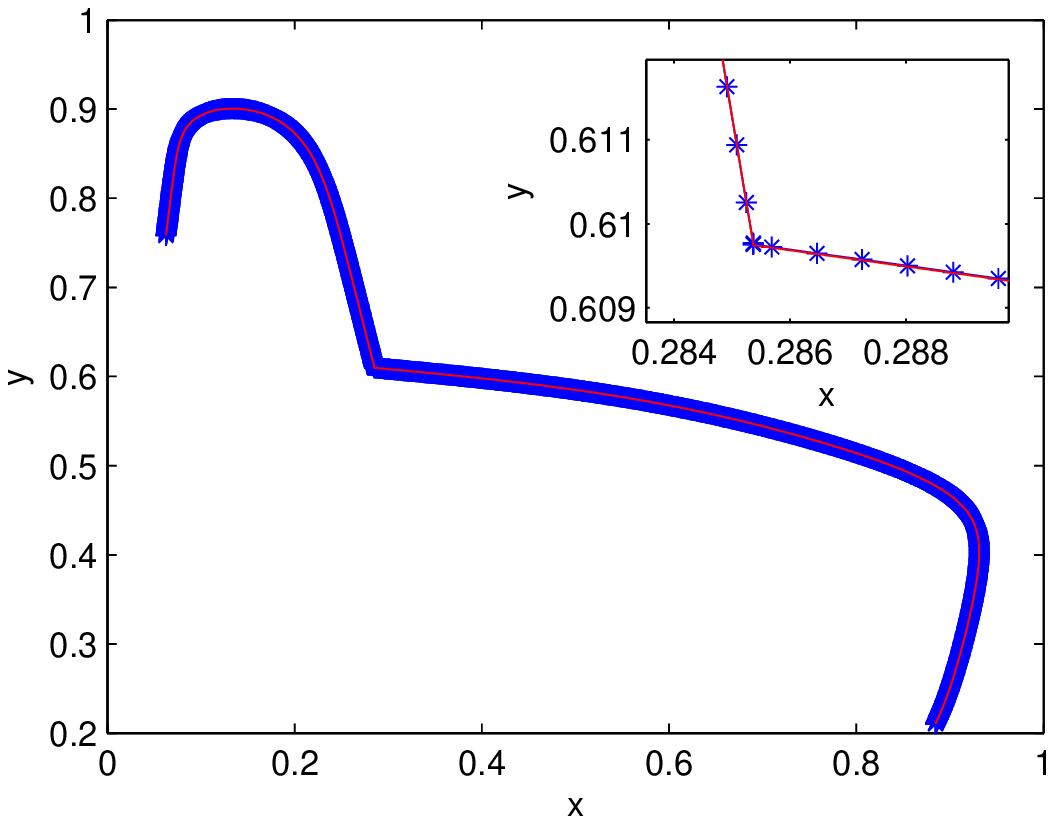}
\end{minipage}
\label{f4b}
}
\caption{The MAM solutions of the paths for the tunnel-diode Model. Solid line: the true solution of minimum action path (when $T$ is optimal) calculated from the gMAM. Curves marked  with ``$*$" : the MAM solutions of the minimum action path. (a) :    $T=30$, the number of points $N=100$  and  (b):  $T=200$, the number of points $N=2000$, respectively.    The inlet is the zoom-in near the saddle. Refer to Section \ref{sec4} for detail    description of this model.  It is observed that only a sufficient large $T$    and a   large number of $N$ can find the satisfying approximation of the true path.
}
\label{f4}
\end{figure}

For problem (\ref{q2}), in many examples, the minimum action path passes through the saddle points with two neighbouring stable states as the starting and ending points. For gradient system, this statement is rigorous \cite{Renthesis}. For non-gradient systems, there is no rigorous statement, since the existence of saddle point is still a question. But for many known examples where some saddle point lies on the separatrix, the optimal path does pass the saddle points \cite{zhouthesis}.  As $\varphi$ crosses a fixed point, it is known that  the time derivative $\varphi_t\rightarrow0$ \cite{zhouthesis,Heymann2006} .  If the original MAM in\mbox{\cite{weinan-MAM2004}}, where the time interval is discretized by equal time step size, is applied,  this property of slowdown in dynamics  near saddle point     brings  the``clustering problem'' to the calculation of the MAP   {\it i.e.}, too many points are around the three critical states while too few points are allocated for shaping the path. The clustering problem increases errors and significantly reduces the accuracy of the algorithm. Another issue for the MAM in  \cite{weinan-MAM2004} is the existence of error for the quasi-potential problem (\ref{q2})  due to the truncation of the time interval. Numerical results show, when $T$ is small, the solution of the MAP   deviates from the true solution, which effectively  {corresponds} to $T=\infty$; when $T$ is   large, the solution of the MAP is close to the true solution (Figure \ref{f4}). Thus a sufficiently  large $T$ is   needed to capture the true MAP.

\subsection{The Adaptive Minimum Action Method}
The adaptive minimum action method was proposed to solve the  clustering problem by numerically adjusting the image points $\{\varphi_i\}_{i=1}^N$ to be equally spaced with respect to arc-length. This method in \cite{aMAM2008} used the simplest discretized form of the integral (\ref{q2}) by the finite difference method and rectangle quadrature
\begin{equation}\label{q21}
\min_{\mbox{
\tiny
$\begin{array}{c}
\varphi_1=\phi_0,\\
\varphi_N=\phi_1\end{array}$
}
} \frac{1}{2}\sum_{i=1}^N|\frac{\varphi_i-\varphi_{i-1}}{\Delta t_i}-b(\frac{\varphi_i+\varphi_{i-1}}{2})|^2\Delta t_i,
\end{equation}
where $\Delta t_i=t_i-t_{i-1}$. Suppose at the $k_{th}$ step of some minimization method for  \eqref{q21}, the image points are not equally spaced in arc-length, then the aMAM has a moving mesh and reparameterization step to redistribute the image points, based on the following relation between $t$ and a new  parameter $\alpha$:
\begin{equation}\label{q7}
\begin{cases}
\frac{d}{dt}\left(\frac{1}{\omega(t)}\frac{d\alpha}{dt}\right)=0, \\
\alpha(0)=0,\ \ \alpha(T)=1,
\end{cases}
\end{equation}
where $\omega$ is the monitor function. Then the problem is actually solved
based on  the uniform partition $\{\alpha_i\}_{i=1}^N=i/N$.

The moving mesh and reparameterization step in \cite{aMAM2008} is implemented as follows: given the partition and image points $(t_i^k,\varphi^k_i),\ i=1,\ldots,N$, at the $k_{th}$ iteration, the aMAM first finds the corresponding partition for the $\alpha$-parameter $\{\alpha_i^k\}_{i=1}^N$ by equation (\ref{q7}), and then finds a new partition $\{\hat{t}_i^k\}$ by interpolating $(\alpha_i^k,t_i^k)$ on the uniform partition $\{\hat{\alpha}_i^k=i/N\}$, and finally locates the new image points $\{\hat{\varphi}^k_i\}_{i=1}^N$ on numerical solution of the path from the new partition $\{\hat{t}^k_i\}_{i=1}^N$ by the cubic spline interpolation.  Then the partition and image points $(\hat{t}_i^k, \{\hat{\varphi}^k_i\}_{i=1}^N)$ are used to solve the numerical scheme of the minimization problem (\ref{q21}) in the next step.

Equation (\ref{q7}) indicates that the partition is denser where the monitor function $\omega(t)$ is larger. Integrate \eqref{q7}, then we can see that
\begin{equation}{\label{q6}}
\frac{\int_{t_i}^{t_{i+1}}\omega(t)dt}{\hat{K}}=\Delta \alpha,
\end{equation}
where $\hat{K}=\int_0^T\omega(t)dt$. Here the grid size $\Delta \alpha$ for $\alpha$ is assumed constant.
 The aMAM in \cite{aMAM2008} selects the monitor function as
\begin{equation}
\label{q14}
\omega(t)=\sqrt{1+\tilde{K}|\varphi_t|^2},
\end{equation}
with some constant $\tilde{K}>0$. Thus by solving equation (\ref{q7}) with a monitor function (\ref{q14}), the aMAM adjusts the partition of time parameter $t$ to be denser where the time derivative $|\varphi_t|$ is larger,  $i.e.$, where we have fast dynamics. And the aMAM usually uses a large enough $\tilde{K}$ to allocate equally spaced image points. We can observe that when $\tilde{K}$ is large, equation (\ref{q6})  indicates that the arc length element, $\int_{t_i}^{t_{i+1}}\tilde{K}|\varphi_t|dt\approx\int_{t_i}^{t_{i+1}}\omega(t)dt =\hat{K}\Delta \alpha$,  is approximately a constant. In this way, the aMAM capturers  the geometric shape of the minimum action path efficiently and thus achieves better accuracy in most cases.

\begin{figure}[!htbp]
\centering
\subfigure[The tunnel-diode Model.]{
\begin{minipage}[b]{0.45\textwidth}
\includegraphics[width=1.06\textwidth]{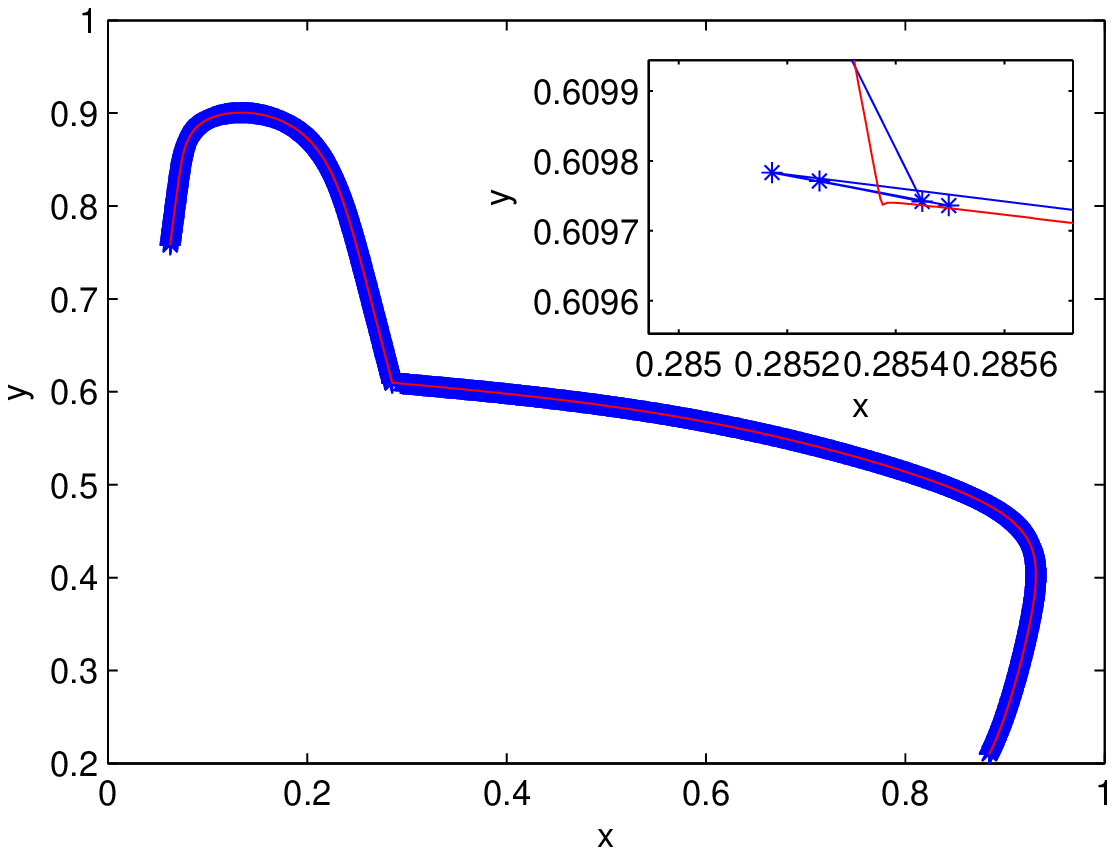}
\end{minipage}
\label{f6a}
}
\subfigure[The Maier-Stein Model.]{
\begin{minipage}[b]{0.45\textwidth}
\includegraphics[width=1.06\textwidth]{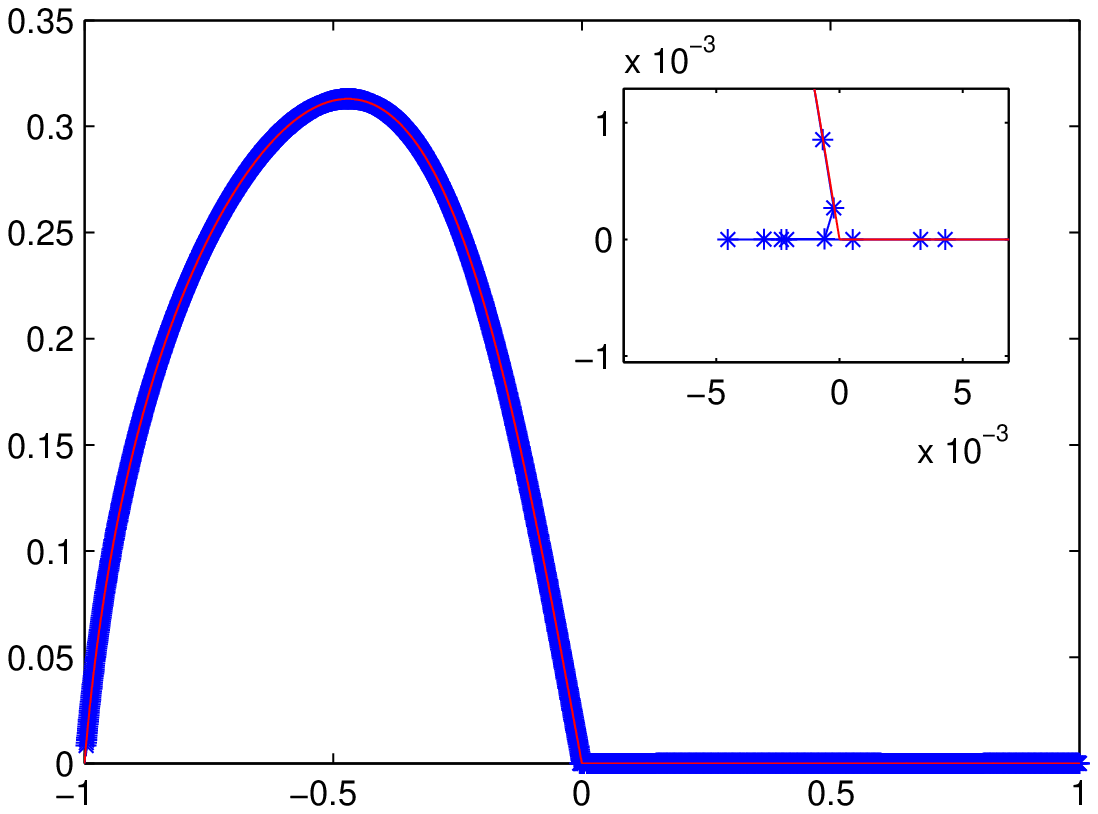}
\end{minipage}
\label{f6b}
}
\caption{The aMAM solutions of the paths for the tunnel-diode model and the Maier-Stein model, with $T=200$ and $T=100$, respectively. Solid line: the true solution of minimum action path (when $T$ is optimal). Curves marked  with ``$*$" : the aMAM solutions of the minimum action path calculated by  the method in \cite{aMAM2008}. The number of points $N=2000$. The inlets are the zoom-in near the saddle point. Refer to Section  \ref{sec4} for the  detail description of these two models.
}
\label{f6}
\end{figure}
However, the singularity of the relationship between the increment of arc length ($d\varphi$) and the time variable ($dt$) can not be resolved, for the fact that $\frac{d\varphi}{dt}\rightarrow0$  at the saddle point is the nature of the minimum action path, particularly for a large $T$ used in practice.  But for a large $T$,  a tangling phenomenon at the saddle may arise numerically in aMAM: the images on the path get locally trapped around the saddle by forming unpredictable zig-zags (the specific  tangling profiles depend on initial condition and the frequency to apply the moving mesh step). And such tangling phenomena only exists in a  local tiny region near the saddle point.  Figure \ref{f6} illustrates this abnormal effect    for  the tunnel-diode model and the Maier-Stein model (Section \ref{sec4}).  In these examples, we use the number of images as large as $N=2000$ for the large $T$ to show that increasing the number of images can not solve the problem. In fact, the more image points used, the more possible for the tangling to occur. It is   noted from numerical experiments  that the above aMAM   in \mbox{\cite{aMAM2008}} is much more likely  to generate such tangling than the original MAM without moving mesh in  \mbox{\cite{weinan-MAM2004}}  (compare Figure \ref{f4b} and \ref{f6a}).

\begin{figure}[!htbp]
\centering
\subfigure[The error.]{
\begin{minipage}[b]{0.45\textwidth}
\includegraphics[width=1.0\textwidth]{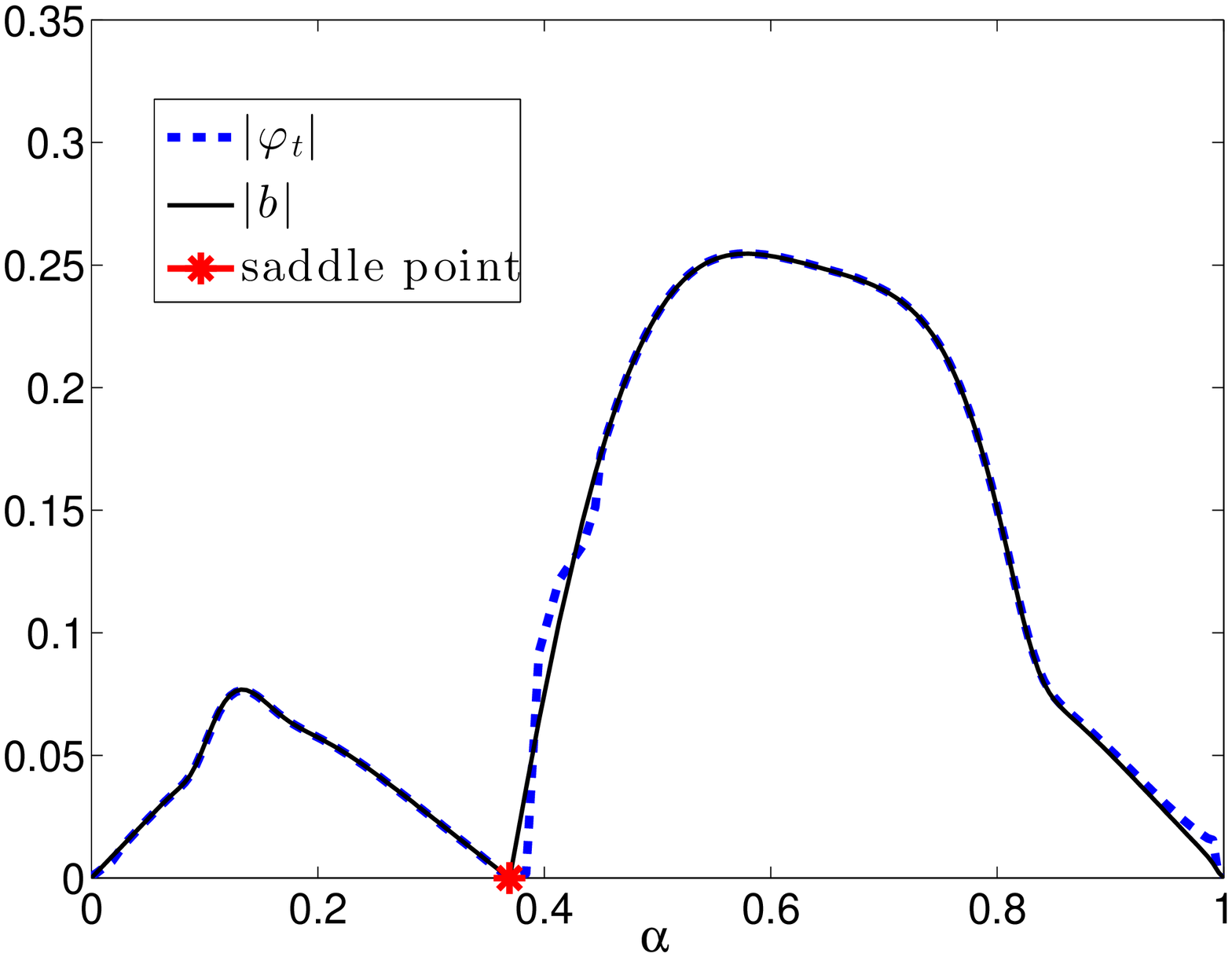}
\end{minipage}
\label{f2a}
}
\subfigure[The magnitude of the gradient $\frac{\delta S}{\delta\varphi}$.]{
\begin{minipage}[b]{0.45\textwidth}
\includegraphics[width=1.0\textwidth]{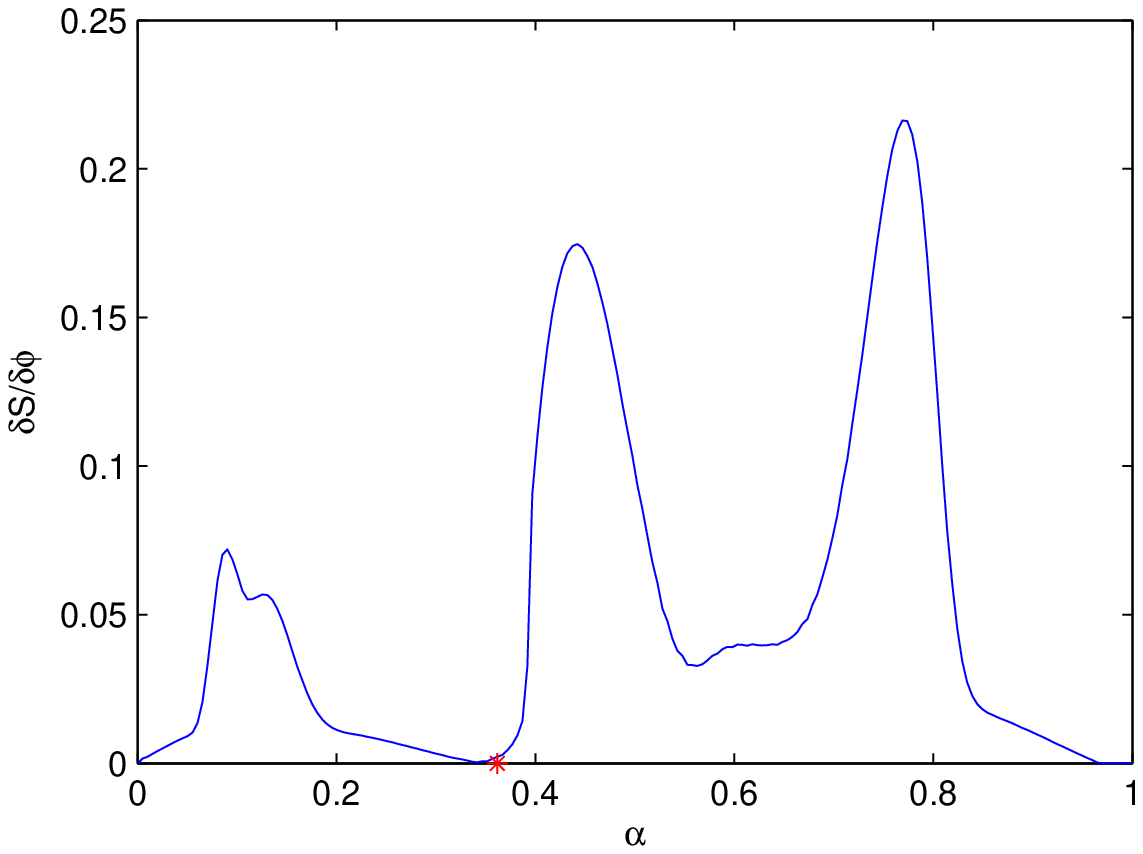}
\end{minipage}
\label{f2b}
}
\caption{(a): The magnitudes of the numerical value of the derivative $\varphi_t$ and the $b(\varphi)$ of the aMAM solutions for the  tunnel-diode  Model.    (b): The magnitude of the gradient $\frac{\delta S}{\delta \varphi}$ of the aMAM solutions for the tunnel-diode Model. The location for the saddle point is  marked by  ``$*$". The number of points used is $N=200$. $T=100$. $\alpha$ is the  curve parametrization. }
\label{f2}
\end{figure}

To illustrate  the potential mechanism for the emergence and persistence of such tangling phenomena in the path optimization algorithms, we show in Figure \ref{f2a} the error in calculating the time derivative $\varphi_t \approx \frac{\varphi_{i+1}-\varphi_i}{\Delta t_i}$ of a typical  aMAM solution after a few numbers of iterations. Theoretically,  $\abs{\varphi_t}=\abs{b(\varphi(t))}$ holds for the true optimal path; refer to \cite{Heymann2006,Stein2016PRE} as well as Section \ref{ssec:31} for this  conclusion. However, our plot clearly indicates that the approximation near the saddle point deviates more from the true value. This observation gives us a strong hint that the use of $ {b(\varphi)}$ is more robust than $\varphi_t$ in our new method in the next section. Actually, in \cite{XWan2016}, the difference quantity $\abs{\varphi_t(\cdot)}-\abs{b(\varphi(\cdot))}$, is exactly the criteria for the posterior error estimate to implement their $h$-adaptivity. Next, we show the magnitude of the gradient of $S$ ({\it i.e.}, the residual of E-L equation) in Figure \ref{f2b}. This figure suggests that near the saddle point, the sensitivity for the {\it objective function} $S$, $\delta S/\delta\varphi$, is quite small, while the error in {\it path} measured by the deviation to the true {solution} is relatively large  near this  saddle  point.  So, any type of local update near saddle point  would not improve the objective function as effective as in other regions. In particular, during the moving mesh step, if  the mesh redistribution or the interpolation of the path from the old partition to the new one accidentally triggers some slight tangling points around the saddle point region, then the gradient $\delta S/\delta\varphi$ does not feel such changes there to effectively untangle the bad points. This  heuristically explains why  in practice, once the tangling happens, it is almost impossible    to  get back in order again. In the original adaptive MAM \mbox{\cite{aMAM2008}},  the reason why it is so easy to  trigger the tangling  is  twofold: (1)  the monitor function \eqref{q14}   used there is not robust  because    it is completely  determined by $\varphi_t$,  and  the  (relative)  numerical error  in  the  approximation of $\varphi_t$  is large at the saddle point. Therefore, the quality of the new partition is not locally very satisfying and deviates from the true arc-length parametrization near saddle points; (2) the cubic spline interpolation is    dangerous: it either smooths out the corner    (if no tangling) or magnify the effect once    one or two image points start to oscillate.  A simple procedure of adding more points locally apparently can not rescue this problem once the distribution of points has been locally ruined.

\section{Methodology}
\label{sec3}
The monitor function of moving mesh strategy in the aMAM scheme is problematic for the fact that it contains the numerical approximation of the derivative $\varphi_t$, $(\varphi_{i}-\varphi_{i-1})/\Delta t_i$. Based on the aMAM scheme, we propose an alternative choice of the monitor function $\omega$ in equation (\ref{q7}) as
 $$\omega=\frac{|b(\varphi)|^{r}}{C}\ ,$$
 where $C=\int_0^T|b(\varphi(t'))|^{r}dt'$ is a normalizing constant, and $r$ is a positive constant. And the relation between the new variable $\alpha$ and time variable $t$ is
\begin{equation}
\label{q12}
\frac{d\alpha}{dt}=\omega(t)=\frac{|b(\varphi(t))|^{r}}{C}.
\end{equation}
With the boundary condition $\alpha(0)=0$, $\alpha(T)=1$, after  integrating equation (\ref{q12}) on $t$, then we have that $\alpha$ is simply
\begin{equation}
\label{q13}
\alpha(t)=\frac{\int_0^t|b(\varphi)|^{r}dt'}{C}\ ,
\end{equation}
and $\alpha\in[0,1]$. For the system modeled by SDE (\ref{q1}), assume $\sigma(\varphi)=I$, $I$ is the identical matrix. The Euler-Lagrange equation associated with the minimization problem (\ref{q2}) is the following boundary value problem (see \cite{weinan-MAM2004})
\begin{equation}
\begin{cases}\label{q5}
\varphi_{tt}- (\nabla b(\varphi)-(\nabla b(\varphi))^T)\varphi_t-(\nabla b(\varphi))^Tb(\varphi)=0,~~~t\in[0,T] \\
\varphi(0)  =  \phi_0,\ \ \ \varphi(T)=\phi_{1}.
\end{cases}
\end{equation}
We can obtain the minimum action path by numerically evolving  the Euler-Lagrange equation  (\ref{q5})
as the steepest descent dynamics of minimizing $S_T(\varphi)$. Denote $\varphi'$ as the derivative of $\varphi$ with respect to $\alpha$, then we can observe that
\begin{equation}\label{q15}
\varphi_t=\omega\varphi',
\end{equation}
\begin{equation}\label{q16}
\varphi_{tt}=\omega^2\varphi''+\omega\omega_{\alpha}\varphi'.
\end{equation}
Plug the two terms (\ref{q15}) and (\ref{q16}) into the Euler-Lagrange equation (\ref{q5}), we have the equivalent form of the Euler-Lagrange equation written in the new variable $\alpha \in [0,1]$:
\begin{equation}\label{q9}
\begin{cases}
0  =  \omega^2\varphi''-\omega (\nabla b(\varphi)-(\nabla b(\varphi))^T-\omega_\alpha)\varphi'-((\nabla b(\varphi))^T)b(\varphi), \\
\varphi(0)  =  \phi_0,\ \ \ \phi(1)=\phi_{1}.
\end{cases}
\end{equation}
The action functional is calculated by
\[
S_T(\varphi)=\frac{1}{2}\int_0^T|\varphi_t-b(\varphi)|^2dt=\frac{1}{2}\int_0^1\frac{1}{\omega}|\omega\varphi'-b(\varphi)|^2d\alpha.
\]

Instead of  solving the Euler-Lagrange equation (\ref{q5}) on time parameter $\{t_i\}_{i=1}^N$  with the moving mesh strategy in equation (\ref{q12}), we are actually solving its equivalent form (\ref{q9}) on  $\{\alpha_i\}_{i=1}^N$.

\subsection{The connection to gMAM}
\label{ssec:31}
In \eqref{q12}, if $r=1$, the parametrization  in $\alpha$ is just the arc-length parametrizaton, when $T$ is optimal ($T=\infty$).
It is proven  in the geometric minimum action method \cite{Heymann2006} and in \cite{Stein2016PRE}  that for the optimal $T$, the solution to minimization problem (\ref{q4}), $\varphi^*$, satisfies the condition that the Hamiltonian $H(\varphi^*,\varphi^*_t-b(\varphi^*))=0$. We then have
\[
0=H(\varphi^*,\varphi^*_t-b(\varphi^*))=\langle b(\varphi^*),\varphi^*_t-b(\varphi^*)\rangle+\frac{1}{2}\langle \varphi^*_t-b(\varphi^*),\varphi^*_t-b(\varphi^*) \rangle
=\frac{1}{2}(|\varphi^*_t|^2-|b(\varphi^*)|^2),
\]
{\it i.e.}, the minimum action path $\varphi^*$ satisfies $|\varphi^*_t|=|b(\varphi^*)|$. Then by $\eqref{q12}$, {\it i.e.}, $\frac{d \alpha}{dt}=\frac{\abs{b(\varphi^*)}^r}{C}$, we have $${\abs{\varphi^*_\alpha}}=  {\abs{\varphi^*_t}}\frac{dt}{d \alpha}= C \cdot {\abs{b(\varphi^*)}^{1-r}} ,$$ where $C=\int_0^T \abs{b(\varphi^*)}^r dt$. This can be understood as an $r$-dependent reparametrization of the path. $r=1$ corresponds to the arc-length parametrization. So $r=1$ gives $\abs{\phi^*_\alpha}\equiv C$, {\it i.e.}, $\alpha$ is the arc-length parameter. Thus, essentially, when $T$ is optimal and $r=1$, our method is equivalent to the gMAM (with the arc length parametrization), but we have more flexibility here. The Euler-Lagrangian equation in the gMAM has the similar form as \eqref{q9} by replacing $\omega(t)$ by $\lambda(\alpha)=\frac{\abs{b(\varphi)}}{\abs{\varphi_\alpha}}$ which involves the derivative of $\varphi$. The monitor function $\omega= {\abs{b(\varphi)}^r} / {\int_0^T\abs{b(\varphi)}^r dt }$ in our method needs the integral in time variable but without any derivatives.

\subsection{The selection of $r$}

We have shown that $r=1$ corresponds to the arc length parametrization. If $r$ is larger than $1$, then the equation \eqref{q12} will favor the region where $\abs{b}>1$ and less points (than the arc-length case $r=1$) will be placed around the saddle point where $\abs{b}\ll 1$. So usually a large $r$ is not recommend to handle the non-smoothness near the saddle points. On the other hand, if $r=0$, then $\alpha=t/T$, this is just the (linearly rescaled) time parameter.

So,  selecting $r$ within $ (0,1]$   seems  a good balance between the arc-length parametrization ($r=1$), which is good for capturing the geometric shape of the whole path, and the physical time parametrization ($r=0$), which  favours the saddle points. Notice that at non-smooth saddle point, we need more image points to improve the accuracy of numerical solution. Using equally spaced image points is no longer a good choice. We can decrease the value of $r$ to make the image points denser at saddle points, but not too dense as for $r=0$. In the following Figure \ref{fig:rsmooth}, we show the profiles of the different parametrizations  with $r=1$ and $r=0.5$, respectively,  of the (same) optimal path  $\varphi(\cdot)=(x(\cdot),y(\cdot))$ for the Maier-Stein model. Note that $r=1$ actually gives the arc length parametrization in $s$. It is  clearly seen that for instance,  the $y$-component $y(s)$ (the dashed line) does not have a continuous derivative $d y/ds $ at the saddle point but if the $\alpha$-parametrization  is associated with   $r=0.5$, both $x(\alpha)$ and $y(\alpha)$ are smooth functions of $\alpha$. In fact, it is observed that the first order derivative $d x/d \alpha$ and $dy/d\alpha$ both vanish at the saddle point for $r=0.5$. The singularity is transferred to the relation between $\alpha$ (for $r=0.5$) and $s$ (for $r=1$): $d\alpha/ ds \propto \abs{b(\varphi)}^{0.5-1}=\abs{b}^{-0.5}$. So in the right panel, at the saddle point where $b=0$, the derivative of $\alpha$ w.r.t. $s$ is infinitely large.
\begin{figure}[htp]
\includegraphics[width=0.48\textwidth]{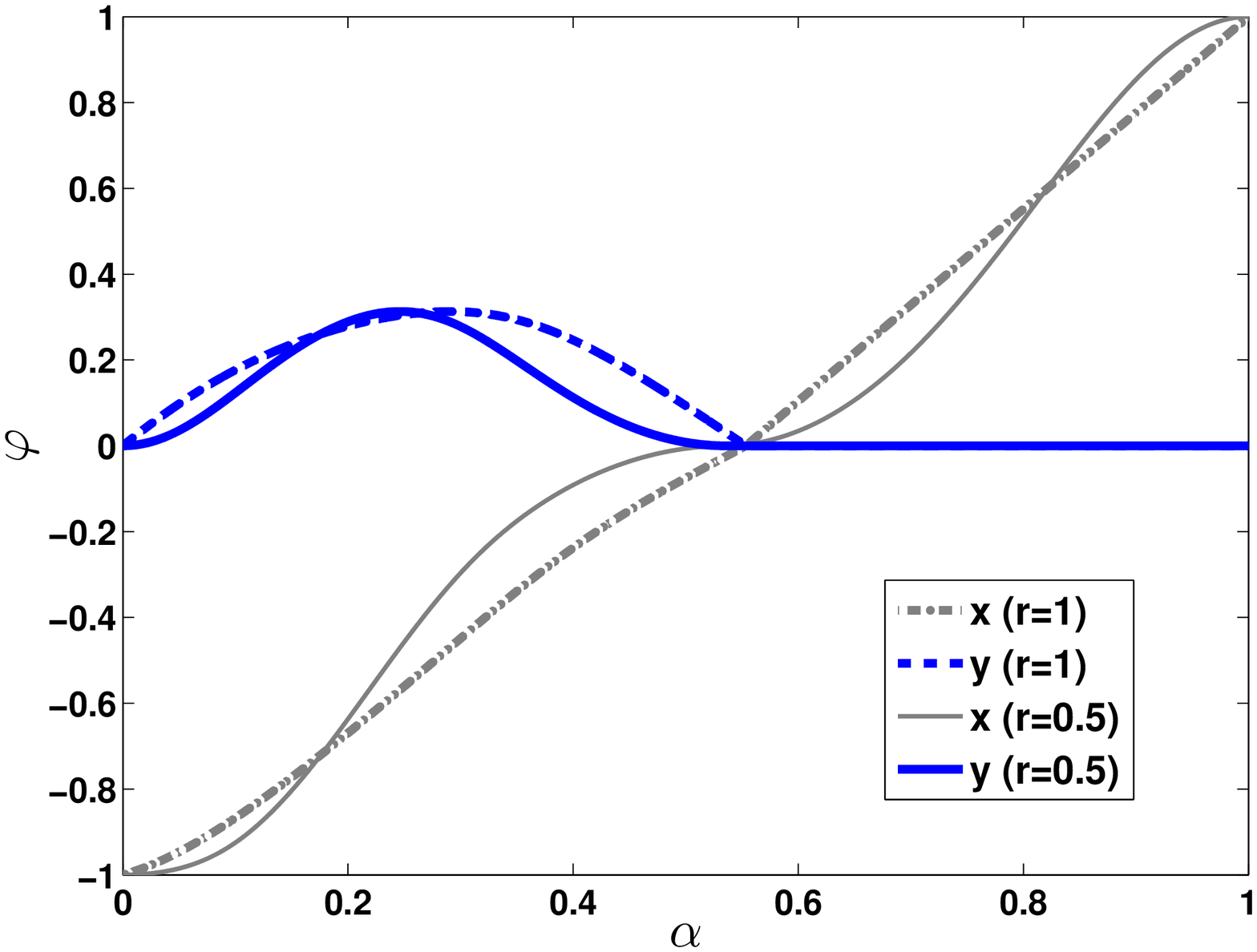}
\includegraphics[width=0.48\textwidth]{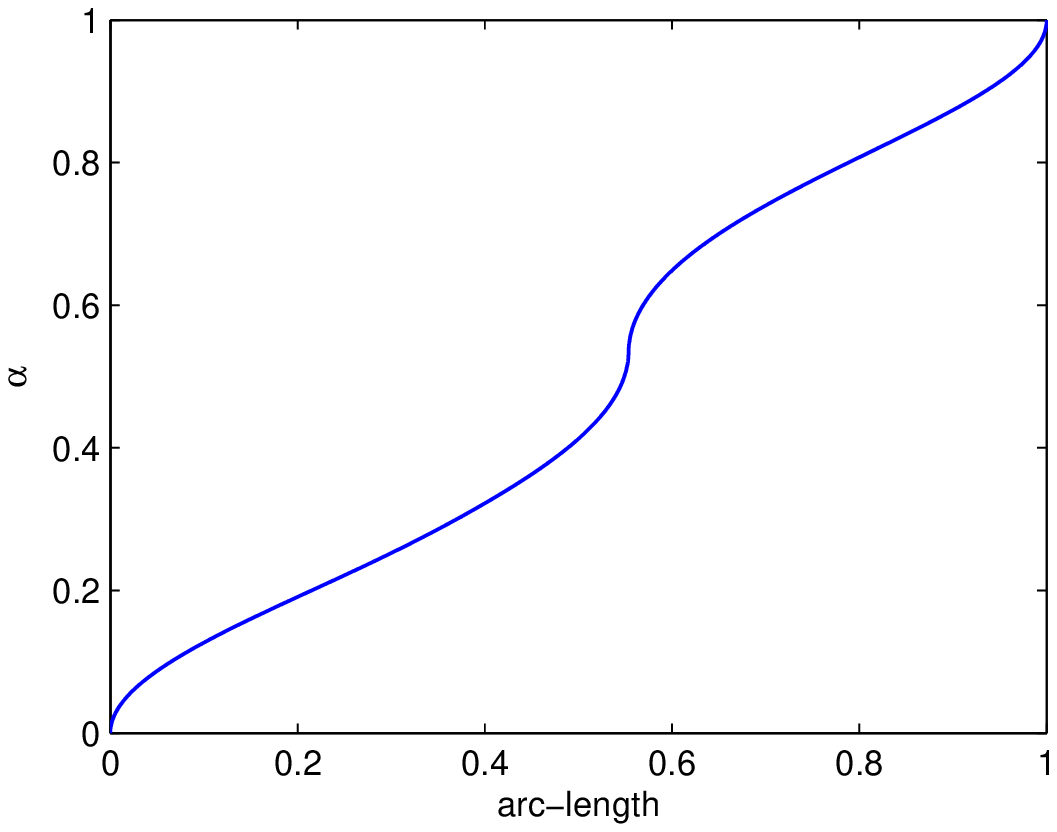}
\caption{The  effect of $r$  on the parametrization of the optimal path  based on \eqref{q12} for  the Maier-Stein model.
Left: The plot of each component in the curve $\varphi(\alpha)=(x(\alpha), y(\alpha))$ v.s. $\alpha$ with different $r=0.5$ and $r=1$.
Note that the two curves of $y$ component for $\alpha>\alpha_*$ with $r=1$ and $r=0.5$ are visually  overlapping.
Right:  The relation between  $\alpha$ with $r=0.5$ and the arc length parameter $s$ ({\it i.e.}, $\alpha$-parameter associated with $r=1$). The saddle point is located at $\alpha_*\approx 0.55$.}
\label{fig:rsmooth}
\end{figure}

Next, we present some heuristic scaling analysis for the observations from the previous figure. Again we use $s$ to refer to the arc-length parameter ($r=1$) and $\alpha$ for the parametrization associated with a general $r>0$. The path is thus referred as $\phi(s)$ or $\varphi(\alpha)=\phi(s(\alpha))$. The non-smoothness in $s$ means at least for one component of the path $\phi$, say the first component $\phi^{1}(s)$, does not have a continuous derivative at $s_*$ which corresponds to the saddle point $\phi(s_*)$, more precisely  $\partial_{s} \phi^1(s_*-)   \neq \partial_{s} \phi^1(s_*+).$  Therefore, for $s$ sufficiently close to $s_*$,
  $$ {\phi^1(s)-\phi^1(s_*)}\propto \abs{s-s_*}^{\gamma},  ~~\gamma\in(0,1],$$
or,
 \[
 \frac{d\phi^1}{ds} (s) ~\propto~ \mbox{sign} (s-s_*)  \abs{s-s_*}^{\gamma-1},
 \]
 where ``$\mbox{sign}$'' is the sign function.
 Note that  we always have the norm of the vector
 $\abs{\phi(s)-\phi(s_*)}\propto   \abs{s-s_*}$ since $\abs{\phi'(s)}\equiv const$.
 Around the saddle point $\phi(s_*)$,
$b(\phi) = b(\phi)-b(\phi(s_*))\propto \abs{\phi(s)-\phi(s_*)}^m
\propto \abs{s-s_*}^m$ and $m=1$ if the Jacobi $\partial_x b(x)$ does not vanish.
By the definition,
 $$d\alpha/dt \propto \abs{b(\varphi)}^r,
 \quad ds/dt \propto \abs{b(\phi)}.$$
 Therefore,
 \[
\begin{split}
 \frac{d\varphi^1}{d\alpha} (\alpha)
& =  ds/d\alpha \times \frac{d\phi^1}{ds} (s)
\\
 & ~  \propto ~\abs{b(\varphi)}^{1-r}  \abs{s-s_*}^{\gamma-1}  \mbox{sign}(s-s_*)
 \\
& ~  \propto ~\abs{s-s_*}^{m-mr+\gamma-1}
 \mbox{sign}(s-s_*)
 \end{split}
 \]
 When $s\to s_*$, $\alpha\to \alpha_*$. So, the equivalent condition for the derivative of the component $\varphi^1$ w.r.t. $\alpha$
 is continuous is that
\[ m-mr+\gamma-1>0, \quad
 {~i.e.}, \quad ~  r < 1 - (1-\gamma)/m.\]
 It easy to see that under this condition, the derivative $d\varphi^1/d\alpha$ actually vanishes at $\alpha=\alpha_*$. For the typical case of $m=1$, the condition becomes
  \begin{equation} \label{eqn:r}
r < \gamma .
 \end{equation}
 $\gamma\leq 1$ describes the type of the singularity of the cusp of the curve in one component. In general case of non-smoothness in multiple components with the corresponding $\set{\gamma_i}$ in each dimension, the condition for $r$ to eliminate the discontinuous
 $d\varphi/d\alpha (\alpha_*)$  is
 \[
 r < \min_i \set{ \gamma_i}.
 \]
In our two numerical examples here (and in most real applications, we guess), $\gamma=1$. So  $r<1$ seems sufficient in theory.
We numerically tested $r=1/2$ and $r=1$, and  find that $r=1/2$ strictly less than one   is indeed   better for the error measured  in $L_\infty$ norm. This observation is consistent with  the empirical conclusion in the recent work \cite{XWan2016}
that the optimal distribution of the finite elements computed from the $h$-adaptivity is not equally spaced in arc-length, but is denser near the saddle point.

\subsection{WENO Scheme for interpolation of the curve $\varphi$}
\label{ssec:WENO}
To improve  the accuracy of interpolations after the new partition is generated, the WENO method \cite{WENO2004JCP} is applied in the reparameterization steps. Next, we briefly present a basic review of this interpolation method. For a function $\varphi(\alpha)$, given the data points $(\alpha_1,\ \varphi_1)$, $(\alpha_2,\ \varphi_2)$, ..., $(\alpha_n,\ \varphi_n)$, $i.e.$, $\varphi_i=\varphi(\alpha_i)$, $i=0,1,2,...,n$, the  Lagrangian interpolation function
$\varphi^*(\alpha)$ is given as an approximation of $\varphi(\alpha)$:
\begin{equation}
\varphi^*(\alpha)=\varphi_0L_{n,0}(\alpha)+\varphi_1L_{n,1}(\alpha)+...+\varphi_nL_{n,n}(\alpha),
\end{equation}
\begin{equation}
L_{n,k}(\alpha)=\frac{(\alpha-\alpha_0)(\alpha-\alpha_1)...(\alpha-\alpha_{k-1})(\alpha-\alpha_{k+1})...(\alpha-\alpha_{n-1})(\alpha-\alpha_n)}{(\alpha_k-\alpha_0)(\alpha_k-\alpha_1)...(\alpha_k-\alpha_{k-1})(\alpha_k-\alpha_{k+1})...(\alpha_k-\alpha_{n-1})(\alpha_k-\alpha_n)}.
\end{equation}
To get the approximation of the function value at $\alpha_*$, where $(\alpha_{i-2}\leq\alpha_*\leq\alpha_{i+2})$, the following three interpolation can be used.
\begin{equation}
\varphi^{(1)}_*=\varphi_{i-2}L_{2,0}(\alpha_*)+\varphi_{i-1}L_{2,1}(\alpha_*)+\varphi_iL_{2,2}(\alpha_*),
\end{equation}
\begin{equation}
\varphi^{(2)}_*=\varphi_{i-1}L_{2,0}(\alpha_*)+\varphi_{i}L_{2,1}(\alpha_*)+\varphi_{i+1}L_{2,2}(\alpha_*),
\end{equation}
\begin{equation}
\varphi^{(3)}_*=\varphi_{i}L_{2,0}(\alpha_*)+\varphi_{i+1}L_{2,1}(\alpha_*)+\varphi_{i+2}L_{2,2}(\alpha_*),
\end{equation}
These three interpolations are built in the local stencils $S_1=\{\alpha_{i-2},\alpha_{i-1},\alpha_{i}\}$, $S_2=\{\alpha_{i-1},\alpha_{i},\alpha_{i+1}\}$, and
$S_3=\{\alpha_{i},\alpha_{i+1},\alpha_{i+2}\}$, respectively. If a larger stencil $S=\{\alpha_{i-2},\alpha_{i-1},\alpha_{i},\alpha_{i+1},\alpha_{i+2}\}$ is used, a fifth-order approximation is then obtained:
\begin{equation}
\tilde{\varphi}_*=\varphi_{i-2}L_{4,0}(\alpha_*)+\varphi_{i-1}L_{4,1}(\alpha_*)+\varphi_{i}L_{4,2}(\alpha_*)+\varphi_{i+1}L_{4,3}(\alpha_*)+\varphi_{i+2}L_{4,4}(\alpha_*).
\end{equation}
Such a high order approximation $\tilde{\varphi}_*$ can be written as a linear convex combination of the three third-order approximations \cite{carlini2005weighted}:
\begin{equation}
\tilde{\varphi}_*=\gamma_1\varphi_*^{(1)}+\gamma_2\varphi_*^{(2)}+\gamma_3\varphi_*^{(3)}.
\end{equation}
The WENO scheme chooses a convex combination of the three approximations $\varphi_*^{(1)},\varphi_*^{(2)},\varphi_*^{(3)}$ as:
\begin{equation}
\tilde{\varphi}_*=\omega_1\varphi_*^{(1)}+\omega_2\varphi_*^{(2)}+\omega_3\varphi_*^{(3)},
\end{equation}
where the weights $\omega_1,\omega_2,\omega_3$ are given based on smooth indicator $\beta_j$ in each stencil $S_j$, $j=1,2,3$.
\begin{equation}
\beta_j=\sum_{l=1}^n(\Delta \alpha)^{2l-1}\int_{\alpha_{i-\frac{1}{2}}}^{\alpha_{i+\frac{1}{2}}}(\frac{d^l}{d\alpha^l}\varphi^{(j)}(\alpha))^2d\alpha.
\end{equation}
 The smoother the function $\varphi^{(j)}(\alpha)$ is in the target cell, the smaller this smoothness indicator $\beta_j$ is. The choice of $\omega_j$ is as follows.
\begin{equation}
\omega_j=\frac{\tilde{\omega}_j}{\tilde{\omega}_1+\tilde{\omega}_2+\tilde{\omega}_3},\ \ \mbox{with} \ \tilde{\omega}_j=\frac{\gamma_j}{(\varepsilon+\beta_j)^2},\  j=1,2,3.
\end{equation}
And in fact the weights might be negative in many cases. Then the above procedures can not provide us
a stable scheme. One can apply the technique in \cite{negativeweights2002JCP} to deal with the negative weights.

\subsection{Numerical Scheme of Improved aMAM for Non-smooth Path}
The main concern of our problem is the case that $T$ is large. For  a large time interval $T$, however, the use of \eqref{q23} in time domain still has some issues near the saddle point for the reason that $\Delta t_i$ is still very large there, although it is  relatively small compared with the total length $T$. So we prefer to working on the $\alpha$-domain $[0,1]$ to solve the Euler-Lagrangian equation \eqref{q9} as a substitution of equation (\ref{q23}). Given the initial path in time discretization, $\varphi_i^k=\varphi(k\Delta\tau,\sum_{j=1}^i\Delta t_j^k)$, $\Delta t_j^k=t^k_j-t^k_{j-1}$, where $k=1$ is the $1_{st}$ iteration, we can find the corresponding   discrete $\alpha$ partition by equation (\ref{q12}). Then we reparametrize this path in  $\alpha$ variable with equal grid size, still denoted as $\varphi_i^k=\varphi(k\Delta\tau,i\Delta \alpha)$ where $\Delta\alpha=1/N$.  We then solve the following numerical scheme of the Euler-Lagrange equation \eqref{q9} written in the variable $\alpha$.
\begin{equation}{\label{q3}}
\begin{cases}
\frac{\tilde{\varphi}_i-\varphi_i^k}{\Delta\tau}  =  (\omega_i^k)^2\frac{\tilde{\varphi}_{i+1}-2\tilde{\varphi}_i+\tilde{\varphi}_{i-1}}{\Delta\alpha^2}+\omega_i^k(\nabla
b(\varphi_i^k)^T-\nabla b(\varphi_i^k)+{(\omega_\alpha)}_i^k){\varphi'}_i^k -\nabla b(\varphi_i^k)^Tb(\varphi_i^k),
\\
\tilde{\varphi}_0 =  \phi_0,\ \ \ \tilde{\varphi}_N=\phi_{1},
\end{cases}
\end{equation}
where $\omega_\alpha$ is calculated by $(\omega_\alpha)_i^k=(\omega_{i+1}^k-\omega_{i-1}^{k})/2\Delta \alpha$. $\omega$ is defined in \eqref{q12} where the constant $C=\int_0^T \abs{b(\varphi)(t)}^r dt$ has to be calculated by using the numerical quadrature from the time partition grids. Both the Euler-Lagrange equation and the WENO scheme we worked on are associated with the parameter $\alpha$, but we   keep updating the discretization with respect to time because it is needed for the computation of the constant $C_k$.

At the ${k}_{th}$ iteration, we need  to numerically calculate the relation between $\alpha$ and $t$ to proceed the reparametrization and moving mesh.  The disrectized form of equation (\ref{q12}) in the moving mesh method is
\begin{equation}
\label{q18}
\Delta \alpha_i^k=\int_{t_i^k}^{t_{i+1}^k}|b(\varphi^k)|^rdt/C_k,
\end{equation}
where $C_k$ is found by calculating the integration $C_k=\int_{0}^{T}|b(\varphi^k)|^rdt$. Notice there is singularity between $\alpha$ and $t$ where $b$ is zero, $i.e.$, when $\alpha_i$ is uniform, $\Delta t_i^k$ can be very large at the saddles where $|b(\varphi_i^k)|\rightarrow0$. Thus the integration (\ref{q18}) should be carefully calculated when $\Delta t_i^k$ is too large, for example, we can interpolate the discrete path with a very fine partition (for instance, break each interval into $M$ pieces) on interval $[t^k_{i-1},t^k_i]$ and then apply the quadrature scheme.

The reparameterization steps are as follows: given the partition and image points $(t_i^k,\varphi^k_i),\ i=1,...,N$ at the $k_{th}$ iteration, we first find the corresponding  parameter $\{\alpha_i^k\}_{i=1}^N$ by equation (\ref{q18}), and then find the new partition $\{\hat{t}_i^k\}$ and the new image points $\{\hat{\varphi}_i\}_{i=1}^N$ on the same path  corresponding to the prescribed uniform partition $\{ {\alpha}_i=i/N\}$. The interpolation between $\alpha$   and the path should be carefully done  by the third-order WENO scheme since the path might be non-smooth.  Then the new   image points  on the uniform $\alpha$-partition   $(  \alpha_i, \{\hat{\varphi}_i\}_{i=1}^N)$  are used to solve the numerical scheme  of the Euler-Lagrange equation \eqref{q3}.

 In summary, we present a new scheme for the calculation of non-smooth minimum action path, for a fixed and large $T$. Given the initial $\{(t_i^0,\ \phi_i^0)\}_{i=0,...,N}$, we update the points as the following algorithm (Algorithm \ref{alg:HgMAM}). In all numerical examples we show here, the initial guess is always the straight line connecting two ends.

\begin{algorithm}[!htbp]
\caption{An improved aMAM algorithm for non-smooth path}
\label{alg:HgMAM}
\begin{algorithmic}[1]
\STATE At the $k_{th}$ iteration, given $(t_i^k,\ \varphi_i^k)$, $i=0,...,N$, calculate $\omega_i^k=\frac{|b(\varphi_i^k)|^{r}}{C_k}$.
\STATE Find $\alpha_i^{k}$ by equation (\ref{q18}).
\STATE Reparameterization: set $\Delta\alpha=1/N$ and  the (fixed) uniform partition  $\alpha_i$, where $\alpha_i-\alpha_{i-1}=\Delta\alpha$. Use
the WENO  scheme to interpolate $(\alpha_i^{k},\ \varphi_i^k)$, and find new image points on $\alpha_i$. Denote the new image points as $\tilde{\varphi}_i^{k}$.
\STATE Given $\alpha_i$, find $t_i^{k+1}$ by interpolating $(\alpha_i^{k},\ t_i^k)$.
\STATE Let $\{\tilde{\varphi}_i\}_{i=1}^{N-1}$ be the solution of (\ref{q10}), and denote $\{\tilde{\varphi}_i\}$ as $\varphi_i^{k+1}$

\begin{equation}
\begin{cases}\label{q10}
\frac{\tilde{\varphi}_i-\tilde{\varphi}_i^{k}}{\Delta\tau}  =  ({\omega_i^k})^2\frac{\tilde{\varphi}_{i+1}-2\tilde{\varphi}_i+\tilde{\varphi}_{i-1}}{\Delta\alpha^2}+\omega_i^k(\nabla
b(\tilde{\varphi}_i^{k})^T-\nabla b(\tilde{\varphi}_i^{k})+{(\omega_\alpha)}_i^k){\tilde{\varphi}'}_i^{k} -\nabla b(\tilde{\varphi}_i^{k})^Tb(\tilde{\varphi}_i^{k}),
\\
\tilde{\varphi}_0 =  \phi_0,\ \ \ \tilde{\varphi}_N=\phi_{1},
\end{cases}
\end{equation}
where ${\tilde{\varphi}'}_i^{k}=(\tilde{\varphi}_{i+1}^k-\tilde{\varphi}_{i-1}^{k})/2\Delta \alpha$, and $\omega_\alpha$ is calculated by
$(\omega_\alpha)_i^k=(\omega_{i+1}^k-\omega_{i-1}^{k})/2\Delta \alpha$.

\STATE Repeat step 1-5 until stoping criterion is fulfilled.

\end{algorithmic}
\end{algorithm}

\section{Numerical examples}
\label{sec4}
\subsection{Example 1. The tunnel-diode model}
Consider random perturbations of the state model of a tunnel-diode circuit.
\begin{equation}
\begin{cases}
\dot{x}=0.5(-h(x)+y)+ \sqrt{\varepsilon}\dot{W}_1 \\
\dot{y}=0.2(-x-1.5y+1.2)+\sqrt{\varepsilon}\dot{W}_2
\end{cases}
\end{equation}
where $h(x)=17.76x-103.79x^2+229.62x^3-226.31x^4+83.72x^5$, and $W_1$ and $W_2$ are independent Wiener processes. There are two stable fix points
$Q_1=(0.0626,0.7582)$ and $Q_2=(0.8844,0.2103)$, and the saddle point is $Q_3=(0.2854,0.6098)$. We use our method to calculate the minimum action path, and compare it with the solution of
the aMAM scheme. The initial path is just a straight line connecting $Q_1$ and $Q_2$.

\subsubsection{Stopping criterion}
We run the iteration algorithm until the difference $$\delta_k:=\max_{i=1}^N |\varphi_i^k-\varphi_i^{k-1}|$$
 is smaller  than a threshold value $\delta(= 10^{-10})$. And we also set the maximum  for the total number of iterations,  denoted by  ``max-it". So, the algorithm stops when either the accuracy threshold  is attained or the maximum number of steps is achieved.

\subsubsection{Real solution}
The numerical results show the solution of the MAP calculated from interval $[0,T]$ is extremely close to the true solution of the MAP at optimal $T=\infty$. Approximately, we take the minimum action path calculated by the gMAM scheme ($i.e.$ the solution of MAP when $T$ is optimal) with 10000 points as real solution. To get the high accuracy at the saddle point, we actually find the exact location of the saddle point first with the Newton method, then use gMAM to calculate the downhill and uphill pieces  separately. Thus, the solution is  the union of these two pieces.

\subsubsection{Error}
To quantify  the error between the numerical path and the true path, we first define the   distance
from a point $x\in \Real^d$ to a continuous path $\varphi=\varphi(\alpha) \subset \Real^d$:
\begin{equation}
\label{def:837}d(x,\varphi)=\min_{z\in\varphi}\abs{x-z}=\min_{\alpha}\abs{x-\varphi(\alpha)},
\end{equation}
where $|\cdot|$ is the standard Euclidian norm in $\Real^d$.
Then the Hausdorff metric between two continuous path $\varphi=\varphi(\alpha)$ and  $\hat{\varphi}=\hat{\varphi}(\beta)$ (they may have different parametrization forms) is
\begin{equation}
d_H(\varphi,\hat{\varphi}):=\max \left\{
\max_{\alpha}\, d(\varphi(\alpha), \hat{\varphi}), ~\max_{\beta}\,d(\hat{\varphi}(\beta),\varphi)\right\}.
\end{equation}
We call this   $L_\infty$ error if $\varphi$ is the numerical solution and $\hat{\varphi}$ is the true solution. This $error_\infty$ describes the largest gap between the the numerical path and the true path. Likewise, to describe the ``averaged'' deviation, we define the $L_2$ error by the line integral in the following sense
\begin{equation}
d_2(\varphi,\hat{\varphi}):= \sqrt{d^2_1/2+d^2_2/2} ,
\mbox{ where~~} d^2_1=
\int_{z\in \varphi }d^2(z, \hat{\varphi})  \,dz,~~
d^2_2= \int_{z\in\hat{\varphi} }d^2(z, \varphi)  \,dz.
\end{equation}
It is clear that these definitions of errors are free of parametrization. If either error is zero, then the
two paths are two identical sets.

In practice, a numerical path $\varphi$ is represented by $m$ images, $\varphi_1,\ \varphi_2,...,\ \varphi_m$, and the true solution obtained from a fine partition is also represented by $K$ images $\hat{\varphi}_j, ~j=1,2,...K$, ($K>m$). Then the $L_\infty$ error in discrete form is
\[
L_\infty (\varphi,\hat{\varphi}):=\max \left\{
\max_{1\leq i\leq m}\, \min_{1\leq j\leq K}
\abs{\varphi_i- \hat{\varphi}_j}, ~\max_{1\leq j \leq K}\,\min_{1\leq i \leq m}\abs{\hat{\varphi}_j -\varphi_i}
\right\}.
\]
And the $L_2$ error in discrete form is
\[
\begin{split}
&L_2(\varphi,\hat{\varphi}):=\sqrt{d_1^2/2+d_2^2/2}, ~~~\quad\mbox{where}
\\
&d_1^2 =\sum_{i=1}^{m-1}
(\min_{1\leq j\leq K} \abs{\varphi_i-\hat{\varphi}_j})^2\abs{\varphi_{i+1}-\varphi_i},
~~~
d_2^2 =\sum_{i=1}^{m-1}
(\min_{1\leq j\leq K} \abs{\hat{\varphi}_i-{\varphi}_j})^2\abs{\hat{\varphi}_{i+1}-\hat{\varphi}_i}.
\end{split}
\]

We first interpolate our numerical solution to $m$ equally spaced points $\varphi_i,\ i=1,2,...m$ by linear interpolation, and compare with the real solution (obtained by the gMAM with an extraordinary large number of points) of $K$ equally spaced points $\hat{\varphi}_j, ~j=1,2,...K$. Here we select $K=m=10000$.

\begin{figure}[!htbp]
\centering
\subfigure[aMAM.]{
\begin{minipage}[b]{0.45\textwidth}
\includegraphics[width=1.06\textwidth]{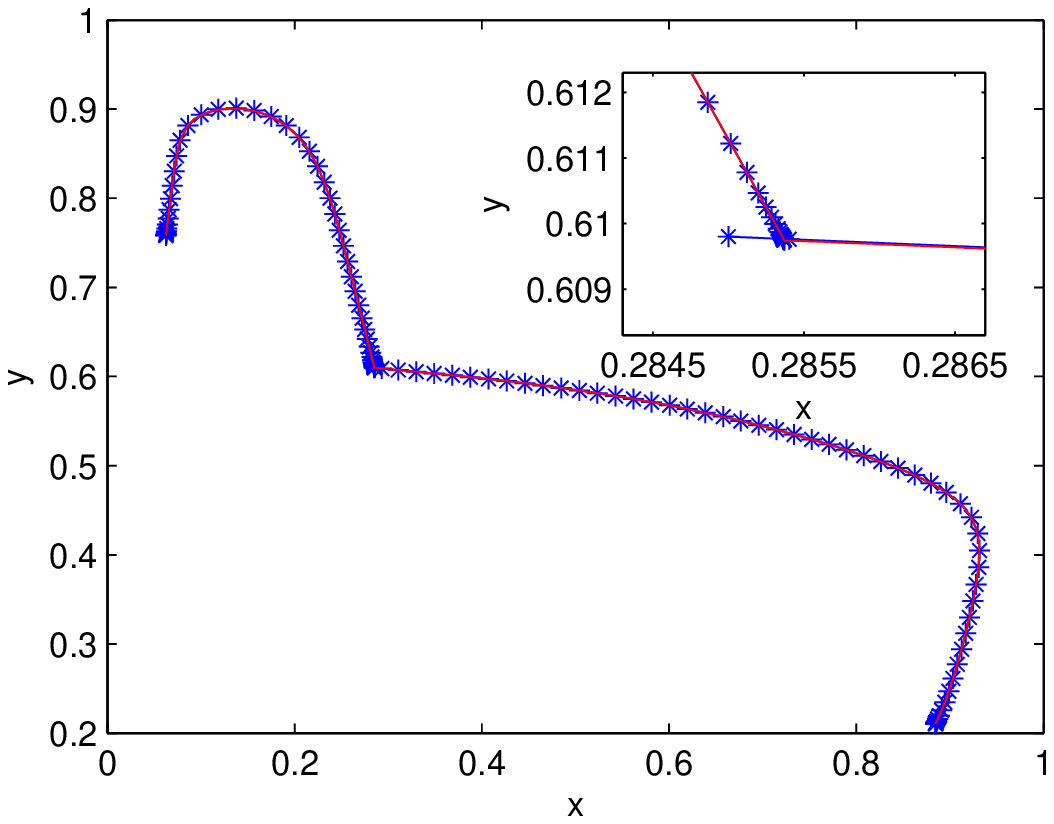}
\end{minipage}
\label{f9a}
}
\subfigure[Our method.]{
\begin{minipage}[b]{0.45\textwidth}
\includegraphics[width=1.06\textwidth]{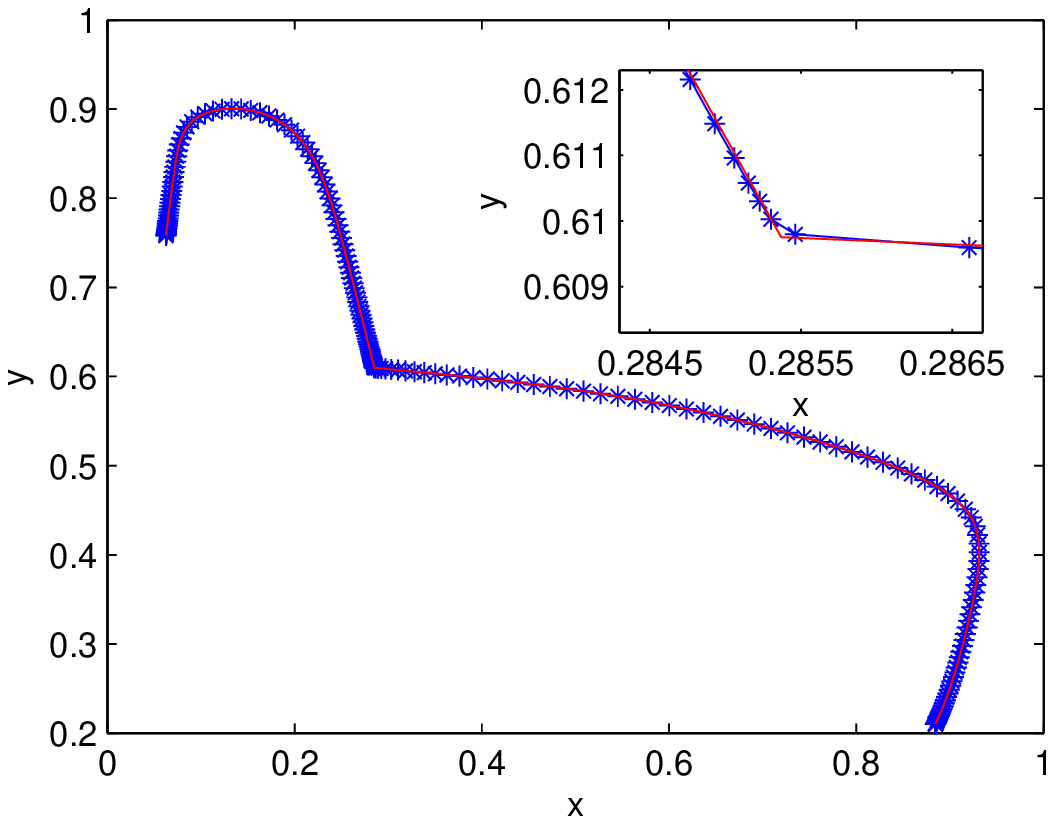}
\end{minipage}
\label{f9b}
}
\caption{Profiles of exact (solid line) and numerical solutions (curves marked  with ``$*$" ) of the tunnel-diode model when $T=200$, the number of points: $N=200$. The number of
iteration: max-it$=1000$. The parameters are selected as $ \Delta \tau=0.1$ and $ r=0.5$. The inlets are  the zoom-in  near the corner where the saddle point is located.}
\label{f9}
\end{figure}

\begin{table}[!htbp]
\begin{center}
\begin{tabular}{c|c|c|c|c}
\hline
\hline
  $N$ & $\delta_k$ & {max-iter} & {$L_2$ Error} & {$L_{\infty}$ Error} \\ \hline
  50  &1.4245e-05& 1000 &2.9000e-03&6.6000e-03\\ \hline
 100  &3.6864e-06& 1000 &4.9028e-04& 1.4000e-03\\ \hline
200  &1.7743e-05& 1000 &1.1779e-04& 2.6127e-04 \\
\hline
\hline
\end{tabular}
\medskip
\caption{$L_2$ error and $L_\infty$ error of our numerical solution when $T=200$. The parameters are $\Delta \tau=0.1$, $r=0.5$. $\delta_k$ is the difference between two iterations when the algorithm stops.}
\label{t4}
\end{center}
\end{table}

\begin{figure}[!htbp]
\centering
\subfigure[$L_2$ error of numerical solutions. ]{
\begin{minipage}[b]{0.45\textwidth}
\includegraphics[width=1.06\textwidth]{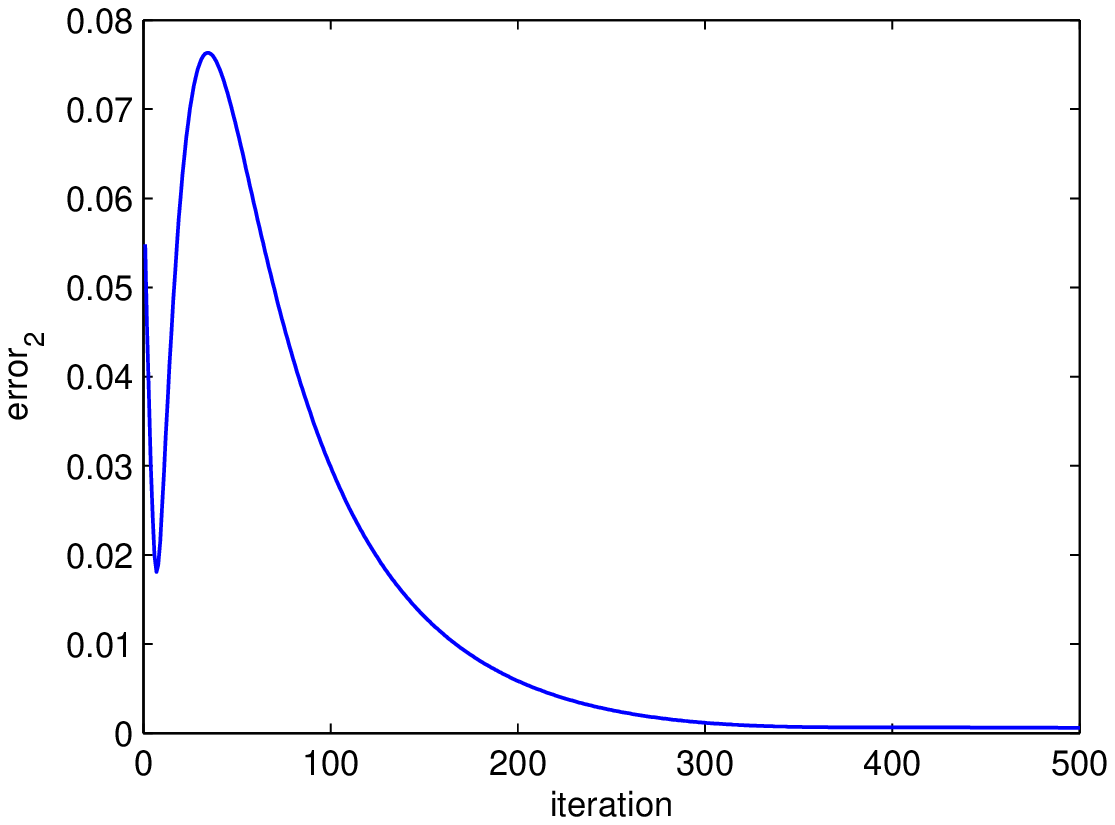}
\end{minipage}
}
\subfigure[$L_\infty$ error of numerical solutions.]{
\begin{minipage}[b]{0.45\textwidth}
\includegraphics[width=1.06\textwidth]{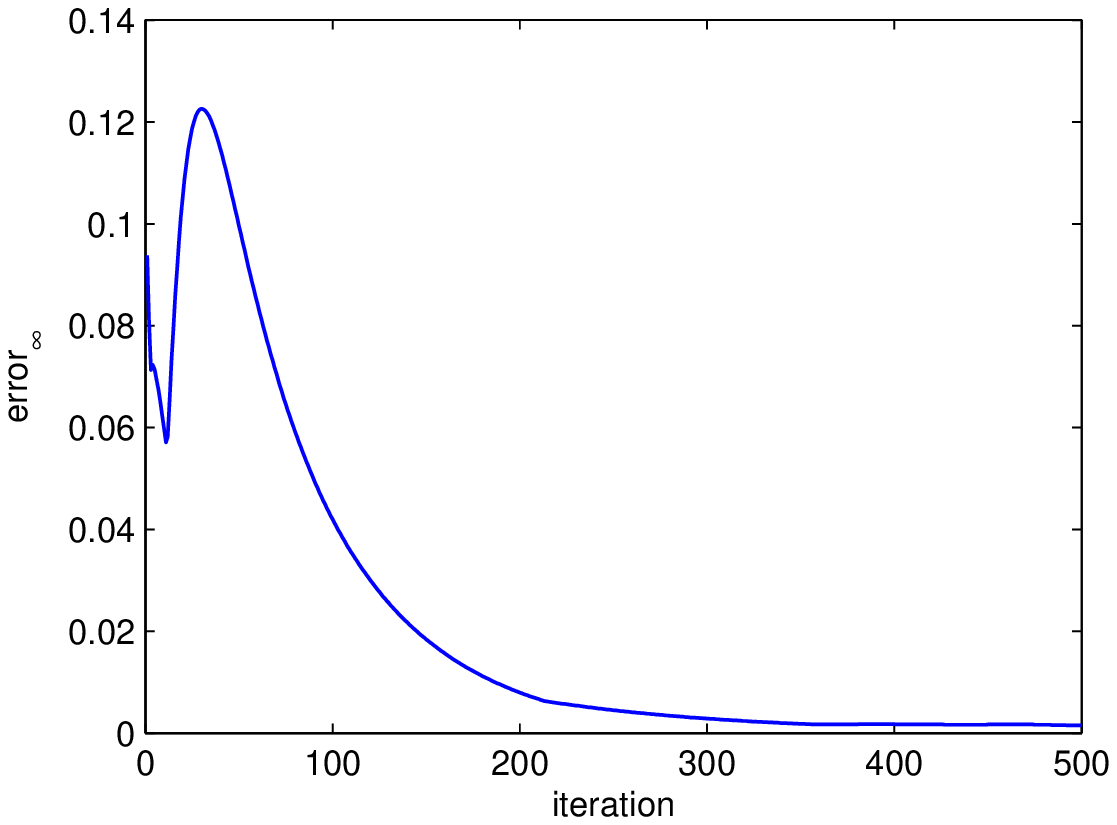}
\end{minipage}
}
\caption{Profiles of error of numerical solutions at each iteration of the tunnel-diode model when $T=200$. The number of points: $N=100$. The parameters are $\Delta \tau=0.1$, $r=0.5$. }
\label{f10}
\end{figure}
\begin{figure}[!htbp]
\centering
\subfigure[$L_2$ error. ]{
\begin{minipage}[b]{0.45\textwidth}
\includegraphics[width=1.06\textwidth]{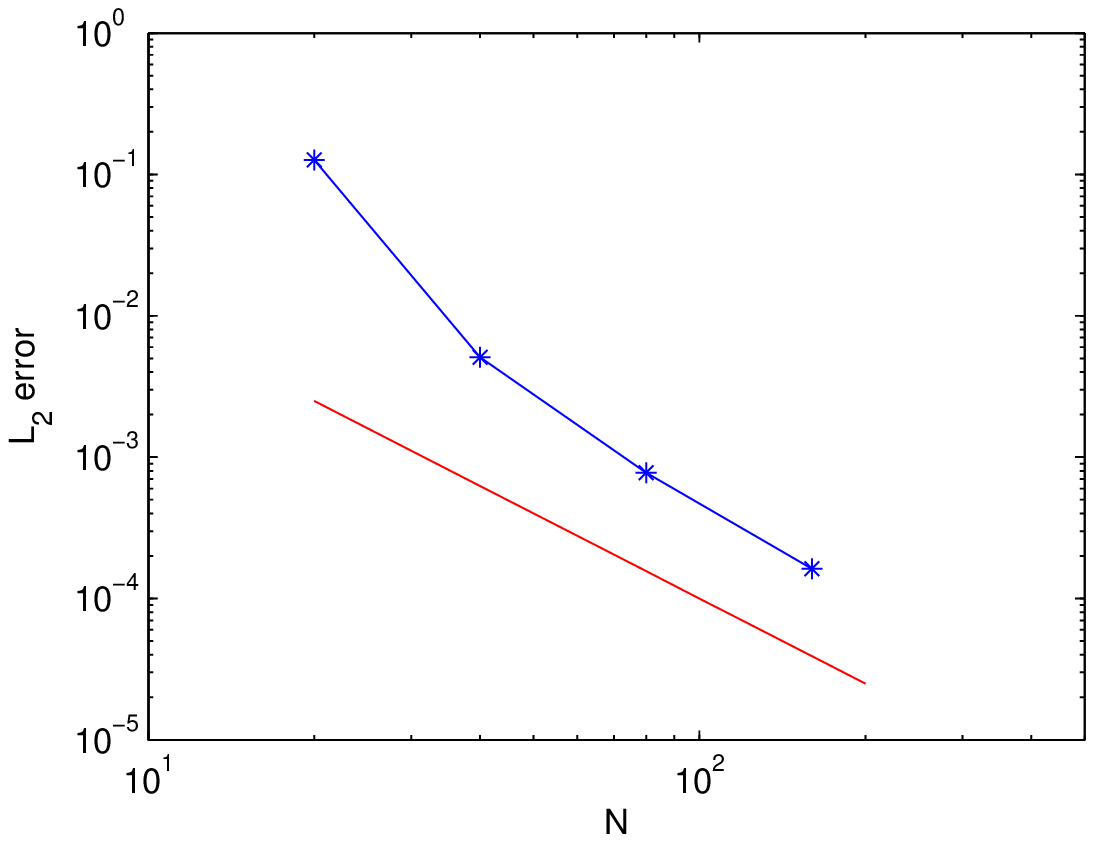}
\end{minipage}
}
\subfigure[$L_\infty$ error. ]{
\begin{minipage}[b]{0.45\textwidth}
\includegraphics[width=1.06\textwidth]{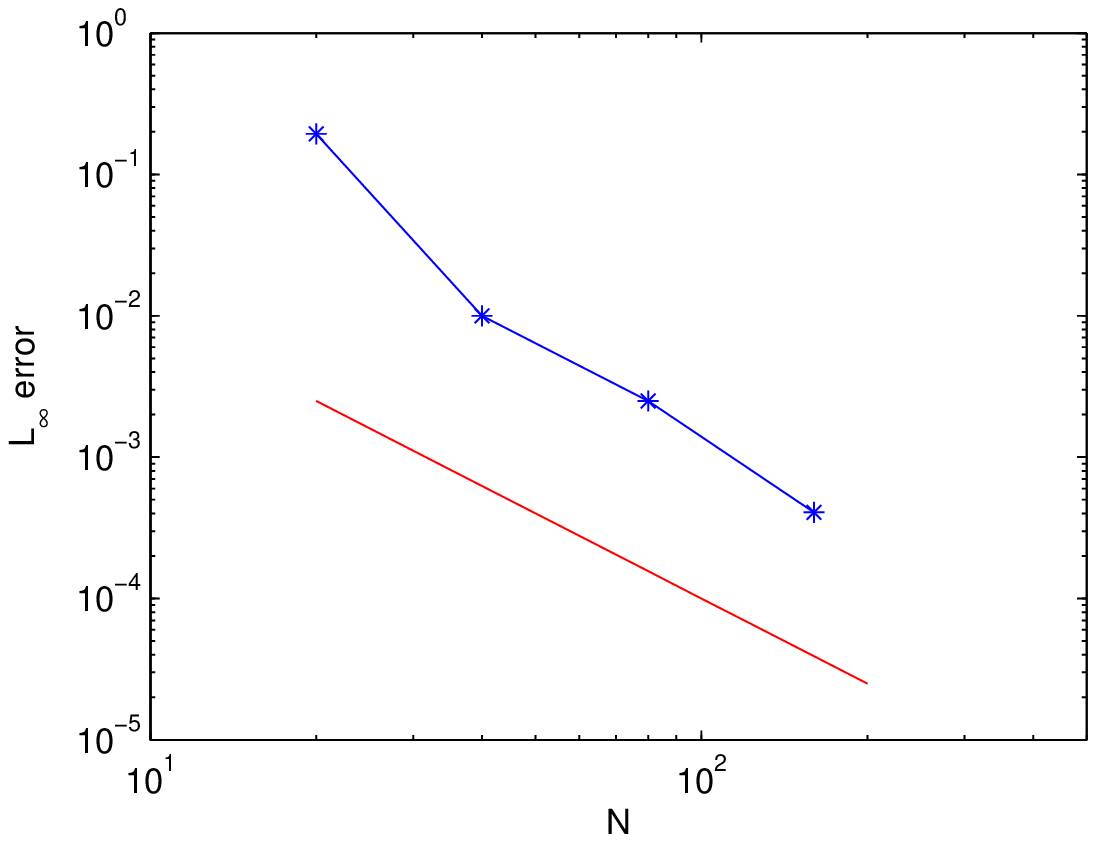}
\end{minipage}
}
\caption{The convergence of   numerical solutions of the tunnel-diode model.  $T=200$, and max-it$=1000$.  The parameters are
 $\Delta \tau=0.1$, $r=0.5$. Curves marked  with ``$*$" : numerical errors of our solution. Solid line: reference line of second order convergence $error \sim N^{-2}$.}
\label{f20}
\end{figure}

In comparison of the numerical results in Figure \ref{f9a} and Figure \ref{f9b}, we find our method improves the efficiency of aMAM at saddles. The numerical illustration shows our method can solve the tangling problem at saddle and directly captures the whole minimum action path. Table \ref{t4} shows $L_2$ errors and $L_{\infty}$ errors of our numerical solution are both small after $1000$ iterations. Figure \ref{f10} shows the $L_2$ errors and $L_\infty$ errors both decay as the number of iteration increases. And our method achieves second order convergence in $L_2$ error and $L_\infty$ error (see Figure \ref{f20}). The above results shows our method performs well for solving the minimum action path.

\subsubsection{Effect of $r$}
\begin{figure}[!htbp]
\centering
\subfigure[$r=0.5$ ]{
\begin{minipage}[b]{0.45\textwidth}
\includegraphics[width=1.06\textwidth]{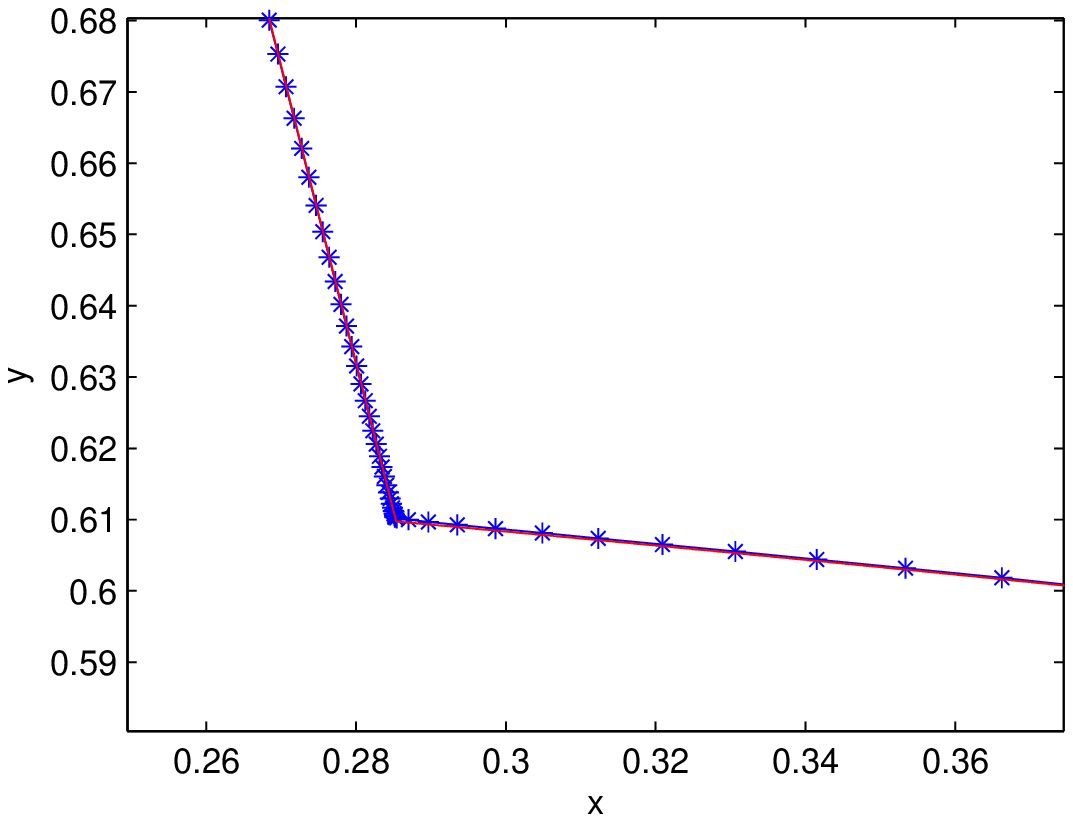}
\end{minipage}
}
\subfigure[$r=1$ ]{
\begin{minipage}[b]{0.45\textwidth}
\includegraphics[width=1.06\textwidth]{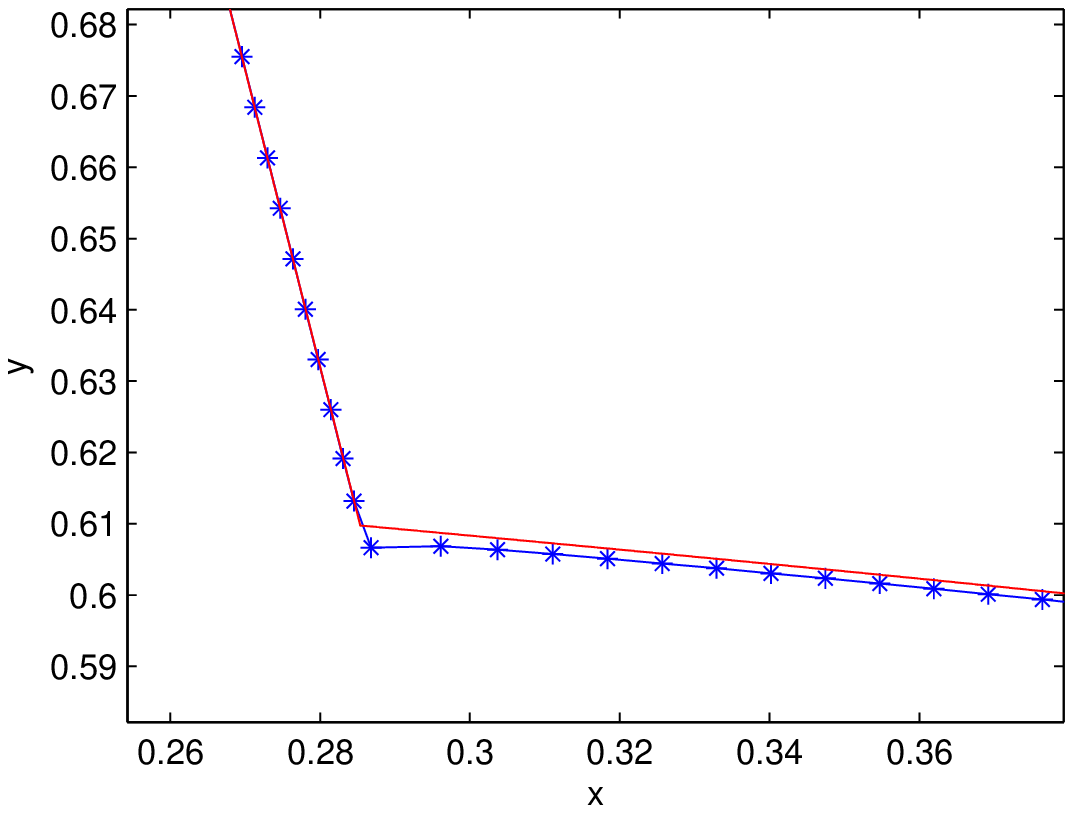}
\end{minipage}
}
\caption{Comparison of  numerical solutions of the tunnel-diode model when $r=0.5$ and $r=1$, zooming at saddle. The number of points: $N=200$. Tolerance: $tol=10^{-6}$. The parameters are selected as $\Delta \tau=0.1$. Solid line: real solution of minimum action path when $T$ is optimal. Curves marked with ``*'': the numerical solution of minimum action path when $T=200$.  }
\label{f21}
\end{figure}

We can compare the result of $r=1$ and $r=0.5$. Figure \ref{f21} and Table \ref{t10} show, the effect of $r$ is to adjust the density of grid points around saddles. When $r=0.5$, the image points are denser at the saddle point than $r=1$, and the numerical solutions are more accurate.

\begin{table}[!htbp]
\begin{center}
\begin{tabular}{c|c|c|c|c}
\hline
\hline
\multirow{2}{*}{$N$}  & \multicolumn{2}{c|} {$r=0.5$} & \multicolumn{2}{c} {$r=1$}\\ \cline{2-5}
  & {$L_2$ Error} & {$L_\infty$ Error} &   {$L_2$ Error} & {$L_\infty$ Error}  \\ \hline
50  & 2.9000e-03& 6.6000e-03 &  2.9000e-03 &4.4000e-03\\ \hline
100   & 4.9028e-04&  1.4000e-03 &  7.3990e-04 &3.3000e-03\\ \hline
200    & 1.1779e-04 & 2.6127e-04 &  5.6424e-04 & 2.4000e-03\\
\hline
\hline
\end{tabular}
\medskip
\caption{$L_2$ error of the numerical solution when $T=200$. The number of points: $N=50,100,200$. Number of iterations: max-it$=1000$. Step size: $\Delta \tau=0.1$.}
\label{t10}
\end{center}
\end{table}
\subsection{Example 2. The Maier-Stein model.}
Consider the following example of a diffusion process proposed by Maier and Stein \cite{Maier1996JSP}:
\begin{equation}
\begin{cases}
\dot{x}=(x-x^3-\beta xy^2)+\sqrt{\varepsilon}\dot{W}_1\\
\dot{y}=-(1+x^2)y+ \sqrt{\varepsilon}\dot{W}_2,
\end{cases}
\end{equation}
where $W_1$ and $W_2$ are independent Wiener processes, and $\beta>0$ is a parameter. For all values of $\beta$ there are two stable equilibrium points $(\pm1,0)$, and an unstable equilibrium point $(0,0)$. And here we select the parameter $\beta=10$. We calculate the minimum action path by our method, and compare it with the solution of the aMAM scheme.  The initial path is simply chosen as $y=-0.5x^2+0.5$. The stopping criterion, real solution, and $L_2$, $L_\infty$ errors are defined as in Section (4.1.1-4.1.3), and $K=m=20000$.

\begin{figure}[!htbp]
\centering
\subfigure[aMAM]{
\begin{minipage}[b]{0.45\textwidth}
\includegraphics[width=1.07\textwidth]{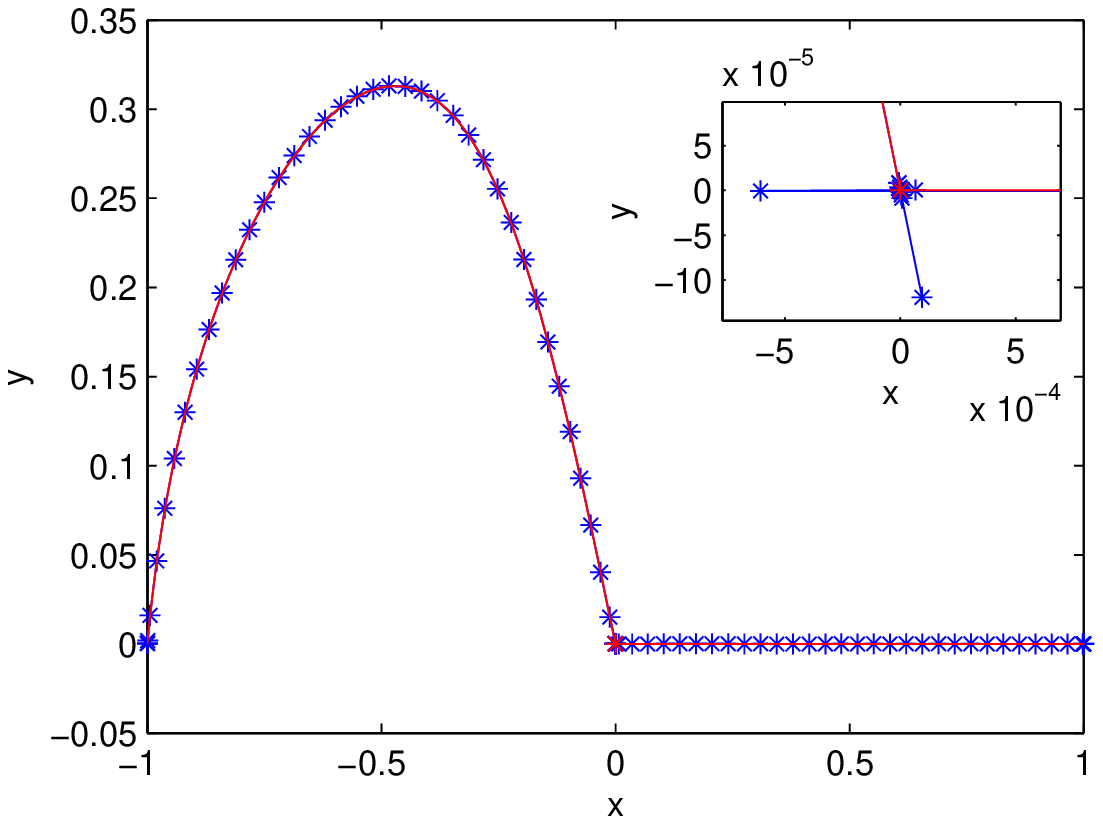}
\end{minipage}
\label{f14a}
}
\subfigure[Our method]{
\begin{minipage}[b]{0.45\textwidth}
\includegraphics[width=1.06\textwidth]{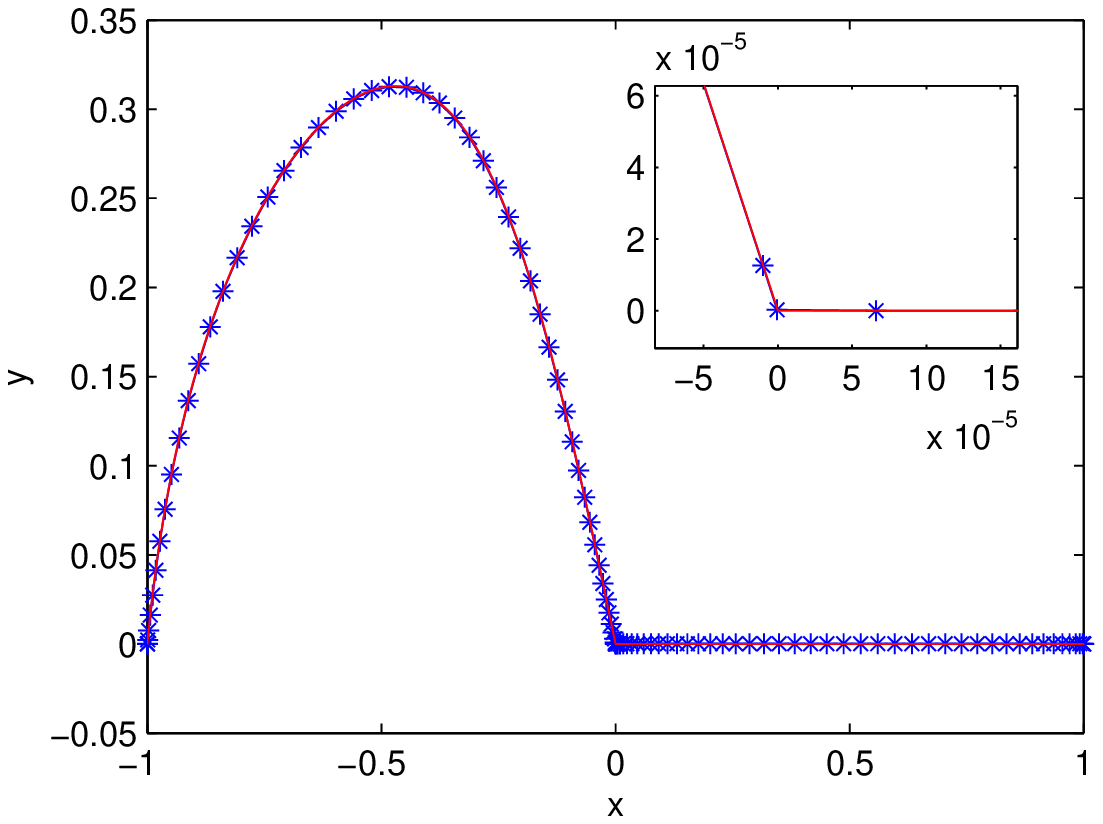}
\end{minipage}
\label{f14b}
}
\caption{Profiles of exact (solid line) and numerical solution (curves marked with ``$*$'') of the Maier-Stein Model when $T=100$, the number of points: $N=100$. Number of iterations: max-it$=1000$. And the parameter are  $\Delta \tau=0.05$, $r=0.5$. The inlets are the zoom-in near the saddle, and there are many image points tangling at saddle in the zoom-in of (a) while there are only three points located in the zoom-in of (b).
}
\label{f14}
\end{figure}

\begin{table}[!htbp]
\begin{center}
\begin{tabular}{c|c|c|c|c}
\hline
\hline
 $N$ & {$\delta_k$} & {max-iter} & {$L_2$ Error} & {$L_{\infty}$ Error} \\ \hline
50  &3.6647e-05& 1000 &7.4775e-04 & 9.0732e-04\\ \hline
100  &3.5343e-06 & 1000 &1.7256e-04 &2.0253e-04 \\ \hline
200  &6.6296e-08 & 1000 &4.7831e-05&5.6775e-05  \\
\hline
\hline
\end{tabular}
\medskip
\caption{$L_2$ error and $L_\infty$ error of our numerical solution when $T=100$. The number of points $N=50,100,200$. The parameters are selected as follows:
$\Delta \tau=0.05$, $r=0.5$.}
\label{t6}
\end{center}
\end{table}

\begin{figure}[!htbp]
\centering
\subfigure[$L_2$ error of numerical solutions. ]{
\begin{minipage}[b]{0.45\textwidth}
\includegraphics[width=1.06\textwidth]{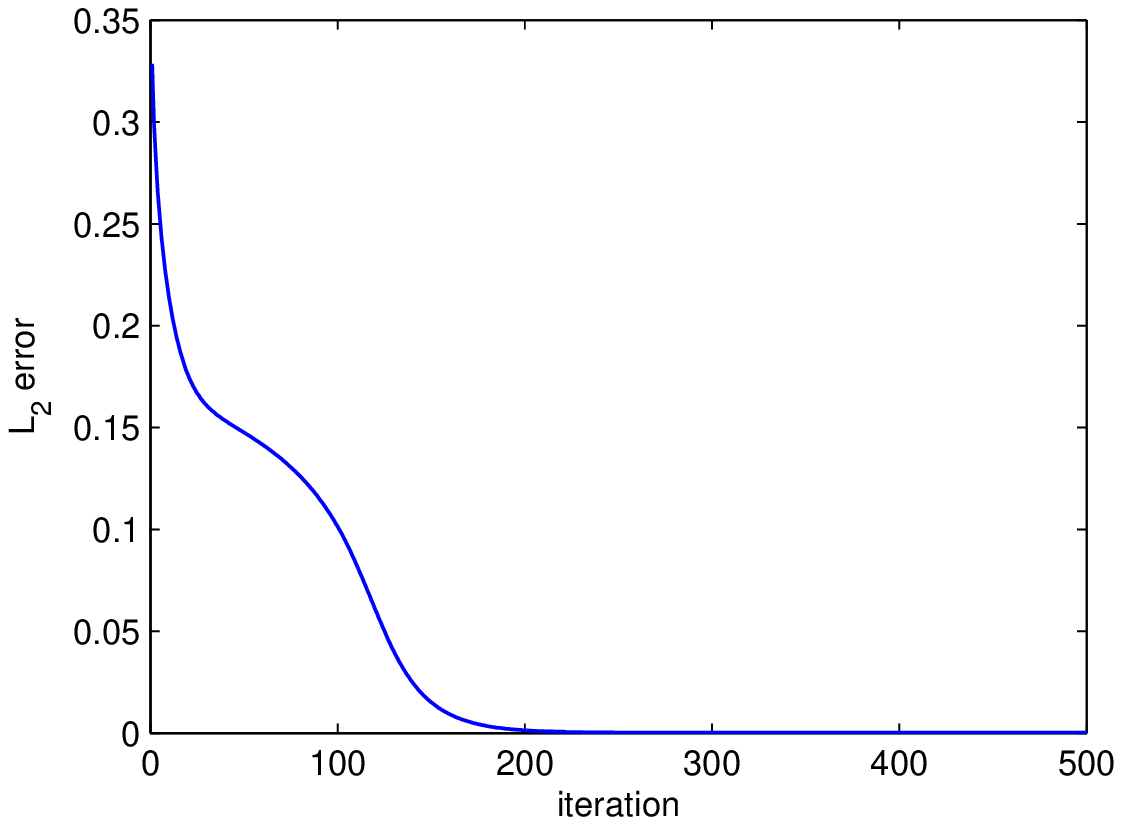}
\end{minipage}
}
\subfigure[$L_\infty$ error of numerical solutions. ]{
\begin{minipage}[b]{0.45\textwidth}
\includegraphics[width=1.06\textwidth]{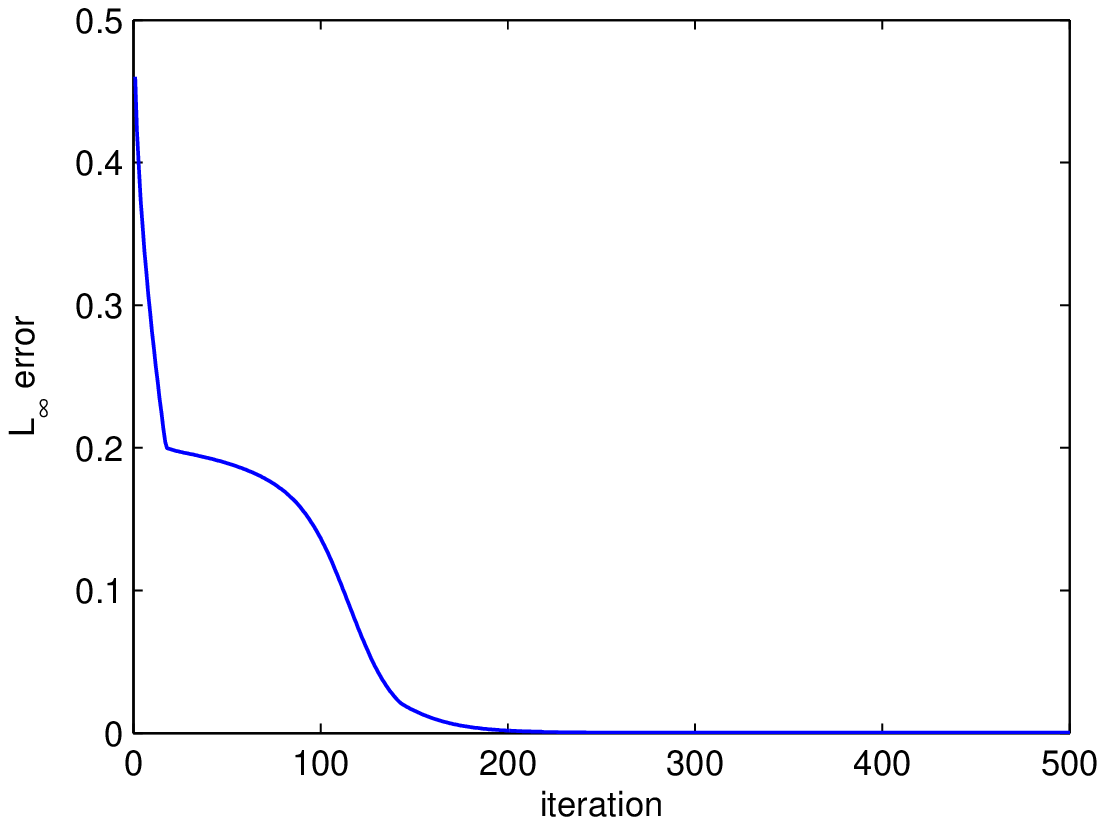}
\end{minipage}
}
\caption{Profiles of error of numerical solutions at each iteration of the Maier-Stein Model when $T=100$. The number of points: $N=100$. The parameters are selected as follows: $\Delta
\tau=0.05$, $r=0.5$.}
\label{f11}
\end{figure}

\begin{figure}[!htbp]
\centering
\subfigure[$L_2$ error. ]{
\begin{minipage}[b]{0.45\textwidth}
\includegraphics[width=1.06\textwidth]{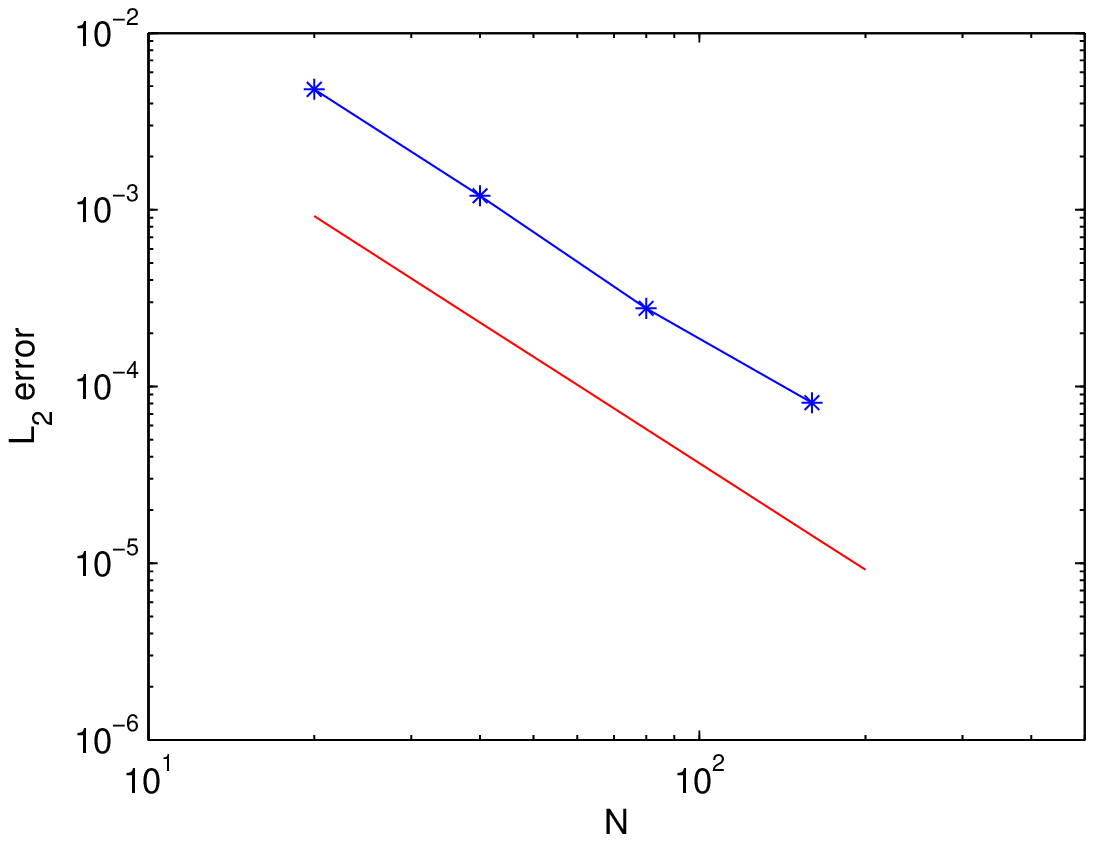}
\end{minipage}
}
\subfigure[$L_\infty$ error. ]{
\begin{minipage}[b]{0.45\textwidth}
\includegraphics[width=1.06\textwidth]{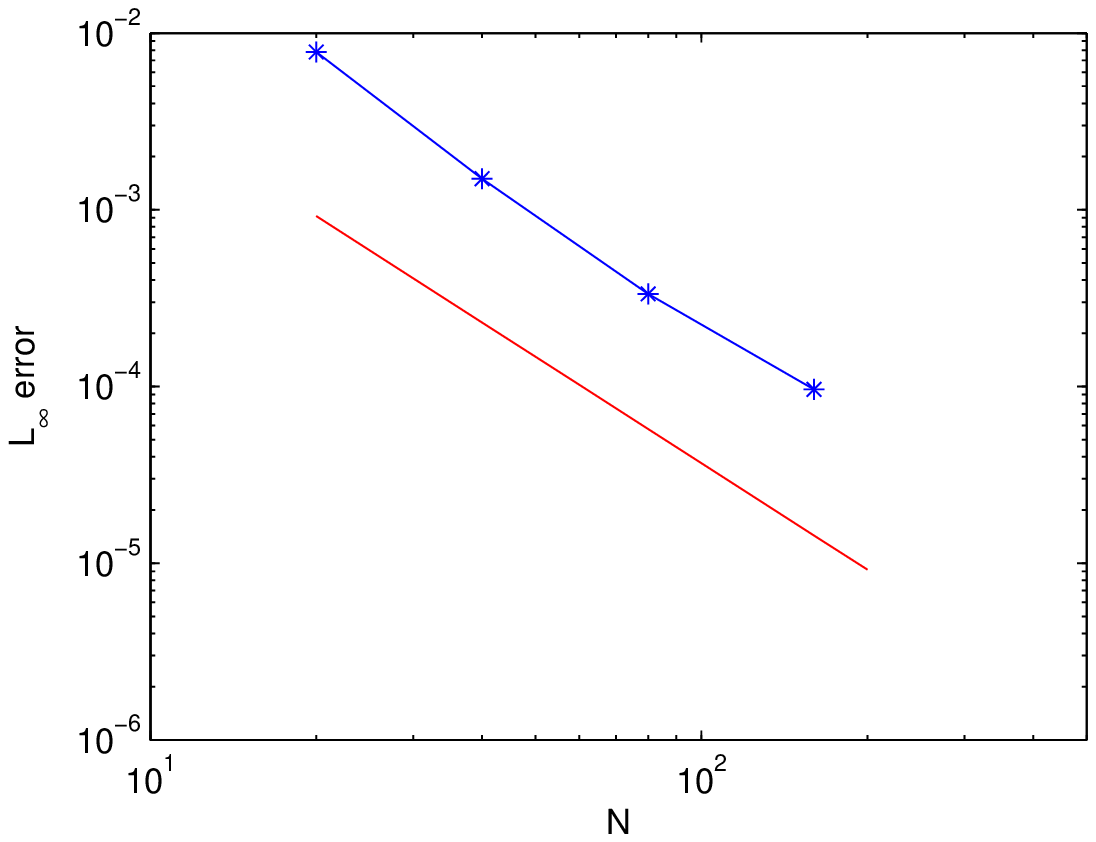}
\end{minipage}
}
\caption{Profiles of error of convergence of numerical solutions of the Maier-Stein Model when $T=100$, with max-it$=1000$, $\Delta \tau=0.05$, and $r=0.5$. Curves marked with ``$*$'': numerical
errors of our solution. Solid line: reference line of second order convergence. }
\label{f19}
\end{figure}

 Figure \ref{f14} shows that our solution captures the whole minimum action path  better than the original aMAM, for our method avoids the tangling problem (Figure \ref{f14a} vs \ref{f14b}). Table \ref{t6} shows $L_2$ error and $L_\infty$ error of our solution are both small after $1000$ iterations. Figure \ref{f11} shows the $L_2$ errors and $L_\infty$ errors of our solution decay at each with the iterations. And Figure \ref{f19} shows second order convergence of our solution in $L_2$ error and $L_\infty$ error.
\subsubsection{Effect of $r$}
\begin{figure}[!htbp]
\centering
\subfigure[$r=0.5$ ]{
\begin{minipage}[b]{0.45\textwidth}
\includegraphics[width=1.06\textwidth]{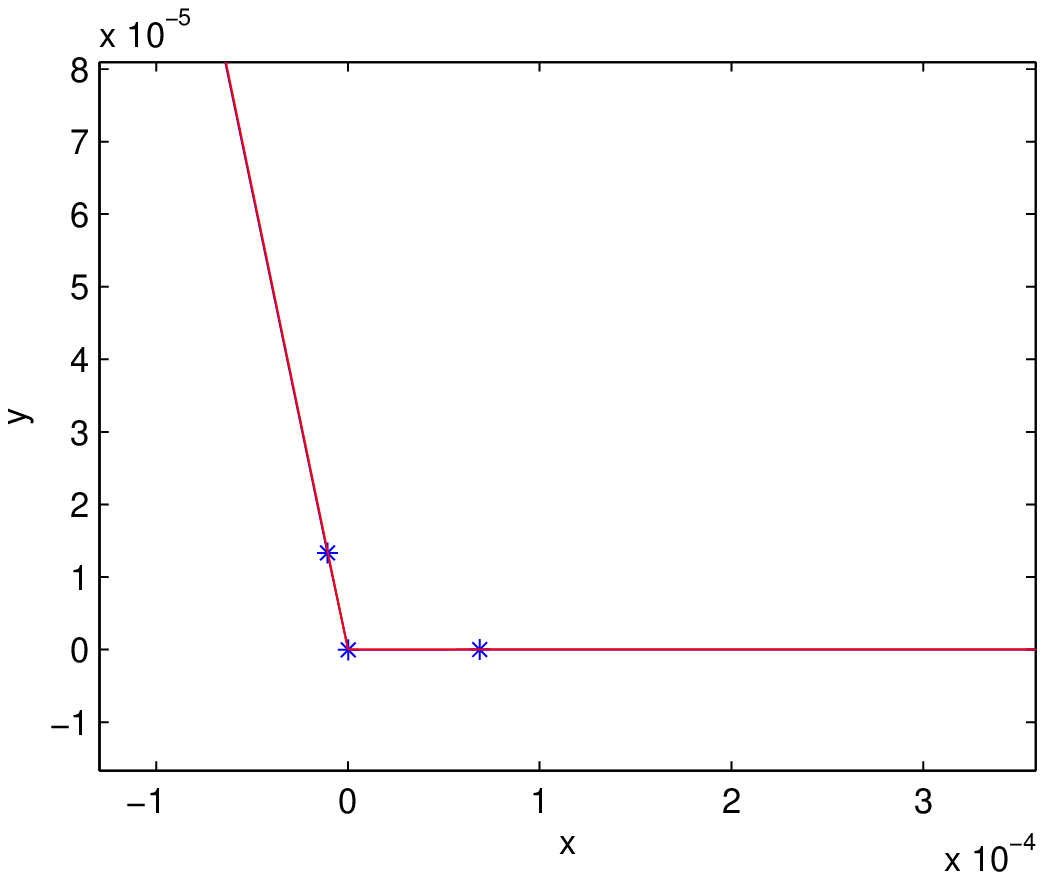}
\end{minipage}
}
\subfigure[$r=1$ ]{
\begin{minipage}[b]{0.45\textwidth}
\includegraphics[width=1.06\textwidth]{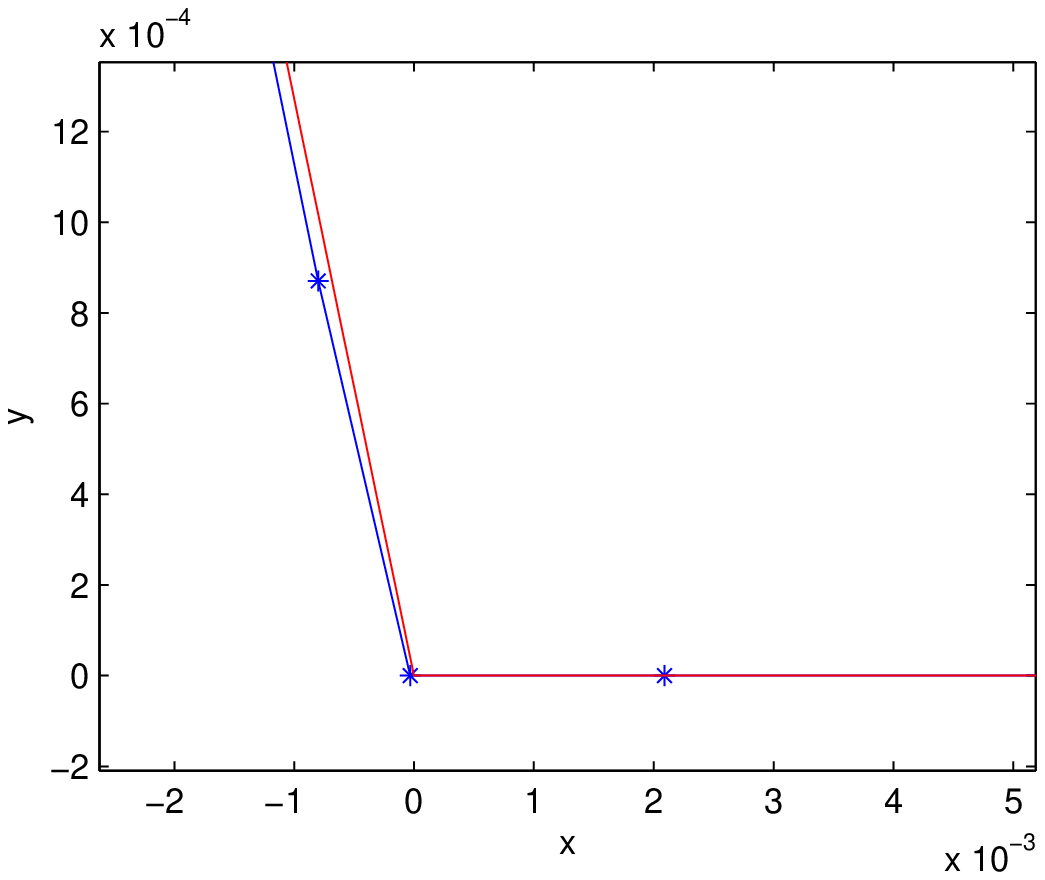}
\end{minipage}
}
\caption{Comparison of  numerical solutions of the Maier-Stein when $r=0.5$, and $r=1$, zooming at saddle. The number of points: $N=100$. The parameters are selected as follows: $ \Delta\tau=0.05$. Number of iterations: max-it$=2000$. Solid line: real solution of the minimum action path when $T$ is optimal. Curves marked with ``$*$'': numerical solution of minimum action
path when $T=100$.}
\label{f23}
\end{figure}

  Also, to illustrate the effect of selecting different values of $r$, we list the result of $r=1$ and $r=0.5$ in Figure \ref{f23} and Table \ref{t11}. From the results, we can see the image points are denser at the saddle point and the numerical errors are smaller when $r=0.5$. This result shows we can adjust the density of grid points by $r$ and thus improve the efficiency of the method.

\begin{table}[!htbp]
\begin{center}
\begin{tabular}{c|c|c|c|c}
\hline
\hline
\multirow{2}{*}{$N$}  & \multicolumn{2}{c|} {$r=0.5$} & \multicolumn{2}{c} {$r=1$}\\ \cline{2-5}
   & {$L_2$ Error} & {$L_\infty$ Error}   &  {$L_2$ Error} & {$L_\infty$ Error}  \\ \hline
50  &7.4775e-04 & 9.0732e-04 &   6.8663e-04 &1.1000e-03\\ \hline
100   &1.7256e-04 &2.0253e-04 &  1.6042e-04 &6.9453e-04\\ \hline
200   &4.7831e-05&5.6775e-05 &  4.1854e-05 &3.2541e-04\\

\hline
\hline
\end{tabular}
\medskip
\caption{$L_2$ error of the numerical solution when $T=100$. The number of points: $N=50,100,200$. Number of iterations: max-it$=1000$. Step size: $\Delta \tau=0.05$. }
\label{t11}
\end{center}
\end{table}

\section{Conclusion}
\label{sec5}
From the numerical results, we can see that our algorithm is efficient in dealing with the path-tangling phenomenon at saddle points. Our scheme improves the original aMAM by excluding the effect of the large numerical errors caused by singularity of the relationship between arc-length and time variable in the calculation of minimum action path. This is mainly achieved with the help of a better choice of the monitor function in moving mesh strategy, which is  free of derivative calculation and has    a   flexible choice of the value $r$. When the value of $r$ is smaller than the  singularity index  $\gamma$, the path  parametrized in our new variable $\alpha$  becomes a  smooth function in $C^1$ so that  the corner problem is avoided.  Our numerical results confirm that  the most efficient parametrization in practical  computation may not be the  arc length parametrization ($r=1$), which  is consist with the recent work  of adaptive mesh refinement based on  the posterior estimate in \cite{XWan2016}. Our idea of using the drift term $b(\varphi)$ can be generalized to the construction of   path parametrized  in other situations like the gMAM or even the string method for some special needs.   Moreover, with the help of WENO in doing reparameterization of non-smooth path, we can find more accurate numerical solutions.   To achieve high order accuracy, one can consider high order numerical schemes of Euler-Lagrange equation (\ref{q9}) and do reparameterization with high order interpolation schemes.

\section*{Acknowledgments}
The research of Xiang Zhou was supported by grants from the Research Grants Council of the Hong Kong Special Administrative Region, China (Project No. CityU 11304314, 11337216, 11304715).


\end{document}